\documentclass[onecolumn,amsmath,amssymb,12pt,superscriptaddress,nofootinbib]{revtex5}
\pdfoutput=1
\usepackage{amsmath}
\usepackage{graphicx}
\usepackage{float}

\begin{document}
	
	\allowdisplaybreaks
	
	\begin{titlepage}
		
		\title{The Wavefunction of Anisotropic Inflationary Universes With No-Boundary Conditions}

		\author{Sebastian F. Bramberger}
		\email[]{sebastian.bramberger@aei.mpg.de}
		\affiliation{Max Planck Institute for Gravitational Physics \\ (Albert Einstein Institute), 14476 Potsdam-Golm, Germany}
		\author{Shane Farnsworth}
		\email[]{shane.farnsworth@aei.mpg.de}
		\affiliation{Max Planck Institute for Gravitational Physics \\ (Albert Einstein Institute), 14476 Potsdam-Golm, Germany}
		\author{Jean-Luc Lehners}
		\email[]{jlehners@aei.mpg.de}
		\affiliation{Max Planck Institute for Gravitational Physics \\ (Albert Einstein Institute), 14476 Potsdam-Golm, Germany}

		\begin{abstract}
			
			\vspace{.3in}
			\noindent
		We study the emergence of anisotropic (Bianchi IX) inflationary universes with no-boundary conditions in the path integral approach to quantum gravity. In contrast to previous work, we find no evidence for any limit to how large the anisotropies can become, although for increasing anisotropies the shape of the instantons becomes significantly different from Hawking's original no-boundary instanton. In all cases an inflationary phase is reached, with the anisotropies decaying away. Larger anisotropies are associated with a much larger imaginary part of the action, implying that the highly anisotropic branches of the wavefunction are heavily suppressed. Interestingly, the presence of anisotropies causes the wavefunction to become classical much more slowly than for isotropic inflationary universes. We derive the associated scaling of the WKB classicality conditions both numerically and analytically.
			\vspace{.3in}
		\end{abstract}
		\maketitle

	\end{titlepage}
	
	\tableofcontents


\section{Introduction}

There is an old, simple idea illustrating the problem of initial conditions: take any state that the universe might be in and evolve it back in time to some primordial epoch -- you will obtain a possible set of ``initial'' conditions for the universe. Considering that our current universe is in many ways special (and certainly non ``generic''), this argument makes it very clear that we will need a theory of initial conditions if we are to understand the history of the universe. The same argument applies to inflationary and ekpyrotic models (for reviews see \cite{Baumann:2009ds} and \cite{Lehners:2008vx}): despite the fact that they are attractors, one may evolve any final state (even arbitrarily non-flat ones) backwards in time to some initial configuration leading to it. Then, without a measure on the set of possible initial configurations, one cannot assess what is likely or unlikely to ensue. What is likely is that a theory of initial conditions will have to be formulated within quantum theory, both because the dynamics near the big bang will require quantum gravity to be truly understood, and simply because we believe that our fundamental laws are quantum mechanical in nature.

In the present paper we will investigate the no-boundary proposal of Hartle and Hawking \cite{Hawking:1981gb,Hartle:1983ai,Hawking:1983hj,Hartle:2008ng}, and the closely related tunnelling prescription of Vilenkin \cite{Vilenkin:1982de,Vilenkin:1983xq,Vilenkin:1984wp,Vilenkin:1986cy}, for anisotropic (Bianchi IX) models of the universe, in the context of an inflationary model. Our focus is on the saddle point geometries that approximate the path integral, and on the classicality of the wavefunction - these are issues that apply equally to both proposals. The inclusion of anisotropies is of interest because it provides the first step in going beyond the often-employed restriction to spatially homogeneous and isotropic minisuperspace models. In that sense these models are already a good deal more realistic than the isotropic ones. Also, it is known that in the approach of a cosmological singularity, the spacetime metric can locally be described to better and better accuracy by precisely a Bianchi IX metric \cite{Belinsky:1970ew}. Thus we may reasonably hope that the Bianchi IX models studied here capture certain salient features of a full superspace analysis. 

Anisotropic models have been studied repeatedly in quantum cosmology, starting from the more qualitative works of Hawking and Luttrell \cite{Hawking:1984wn}, and Moss and Wright \cite{Wright:1984wm}. Various approximate solutions to the Wheeler-DeWitt equation were given by Del Campo and Vilenkin \cite{delCampo:1989hy}, by Amsterdamski \cite{Amsterdamski:1985qu} and by Duncan and Jensen \cite{Duncan:1988zq}. These works provided valuable first insights into the existence and properties of anisotropic instantons. More recently, Fujio and Futamase instigated a more systematic numerical study, in which they found an obstruction to constructing instantons with large anisotropies \cite{Fujio:2009my}. 

Here we wish to extend these studies. We will show that Bianchi IX instantons satisfying the no-boundary regularity conditions may actually be constructed with arbitrary anisotropies. A non-trivial feature is however that care must be taken in choosing a contour of integration in the complex time plane, as for increasing anisotropies singularities start to appear, and the standard contour (originally employed by Hawking since the earliest works \cite{Hawking:1981gb}) becomes inappropriate. The visual methods developed in \cite{Battarra:2014kga,Battarra:2014xoa,Bramberger:2016yog} are well suited to reveal this feature, and readily suggest better contours. 

Even though we do not find any limit to how large the anisotropies can be at a given instant, all classical histories implied by the instantons undergo inflationary dynamics, just as is the case for isotropic models \cite{Hartle:2008ng}, and thus the anisotropies are quickly diluted away. Nevertheless, we find an interesting effect induced by the anisotropies: they cause the wavefunction of the universe to become classical, in a WKB sense, more slowly than in the isotropic case. More specifically, isotropic inflationary universes satisfy the WKB conditions (that the amplitude of the wavefunction should vary slowly compared to the phase) approximately in inverse proportion to the amount of volume created, while anisotropic universes do so only in inverse proportion to the linear size of the universe. We show this result numerically, and prove it analytically for constant equation of state.

\section{The Anisotropic Minisuperspace Model}

We will consider a model of gravity coupled to a scalar field $\phi$ moving in a potential $V(\phi),$
\begin{eqnarray}
S &=& \int d ^4 x  \sqrt{-g} \left( \frac{R}{2} - \frac{1}{2} g ^{\mu \nu} \partial _{\mu} \phi\, \partial _{\nu} \phi - V( \phi) \right) \,,\label{LorentzianAction}
\end{eqnarray}
where we are using natural units $8\pi G=c=\hbar=1.$ We are interested in studying the effects of anisotropies, and to this effect we choose the spacetime metric to be of Bianchi IX form,
\begin{align}
ds_{IX}^2 = - N^2(t)dt^2 + \sum_m \left( \frac{l_m(t)}{2} \right)^2 \sigma_m^2\,,
\end{align}
where $\sigma_1 = \sin\psi d\theta - \cos \psi \sin \theta d\varphi$, $\sigma_2 = \cos \psi d\theta + \sin \psi \sin \theta d \varphi$, and $\sigma_3 = - (d\psi + \cos\theta d\varphi)$ are differential forms on the three sphere such that $0 \leq \psi \leq 4 \pi$, $0 \leq \theta \leq \pi$, and $0 \leq \phi \leq 2 \pi.$ It is particularly useful to re-write the three scale factors as (we will employ the original definition of Misner \cite{Misner:1969hg} -- note that some authors re-scale the $\beta$s by a factor of $2$),
\begin{align}
l_1(t) &= a(t) \exp \left(\frac{1}{2}\left(\beta_+(t) + \sqrt{3}\beta_-(t)\right)\right) \\
l_2(t) &= a(t) \exp \left(\frac{1}{2}\left(\beta_+(t) - \sqrt{3}\beta_-(t)\right)\right) \\
l_3(t) &= a(t) \exp \left(-\beta_+(t)\right)
\end{align}
which makes it obvious that $a$ will yield information about volume change while the $\beta$s quantify shape change. When $\beta_- = \beta_+ = 0$ one recovers the isotropic case. The Lorentzian action in these coordinates becomes
\begin{align}
S = 2\pi^2 \int dt N a \left[  \frac{1}{N^2}\left( -3\dot{a}^2 +a^2 \left(\frac{1}{2}\dot{\phi}^2 + \frac{3}{4}\dot{\beta}^2_+ + \frac{3}{4}\dot{\beta}^2_- \right) \right) - \left( a^2  V(\phi) + U(\beta_+, \beta_-)\right)\right]\,,
\end{align}
where 
\begin{align} \label{anisotropypotential}
U(\beta_+, \beta_-)  = - 2 \left( e^{ 2 \beta_+ } + e^{-\beta_+ - \sqrt{3}\beta_-} + e^{-\beta_+ + \sqrt{3}\beta_-} \right) + \left( e^{ -4 \beta_+ } + e^{2\beta_+ - 2\sqrt{3}\beta_-} + e^{2\beta_+ + 2\sqrt{3}\beta_-} \right)\,.
\end{align}

\begin{figure}[h] 
\begin{center}
\includegraphics[width=0.6\textwidth]{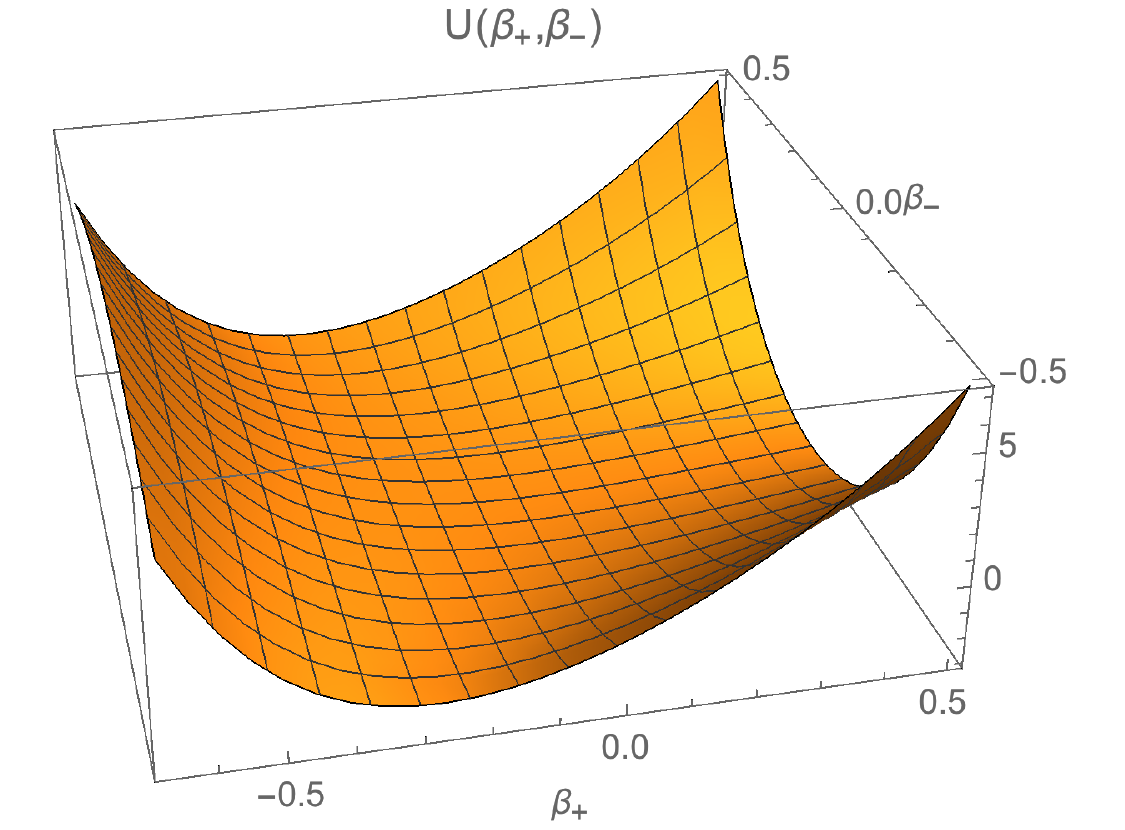}
\caption{A plot of the anisotropy potential $U(\beta_+,\beta_-)$. The minimum is at $U(0,0)=-3.$ Around the minimum the potential has an approximate circular symmetry, which at larger values of the anisotropy parameters morphs into the symmetry of an equilateral triangle.}
\label{fig:U}
\end{center}
\end{figure}

Varying with respect to the lapse $N$ we obtain the Friedman constraint equation
\begin{align} \label{Friedman}
3\dot{a}^2 =a^2 \left(\frac{1}{2}\dot{\phi}^2 + \frac{3}{4}\dot{\beta}^2_+ + \frac{3}{4}\dot{\beta}^2_- \right)  +  N^2 \left( a^2 V(\phi) + U(\beta_+, \beta_-)\right)\,,
\end{align}
while the equations of motion for $a,\beta_+,\beta_-$ are given by
\begin{align}
& \frac{\ddot{a}}{a} + \frac{1}{2}\frac{\dot{a}^2}{a^2} - \frac{2}{aN}\dot{a}\dot{N} + \frac{3}{8}\left( \dot{\beta}^2_+ + \dot{\beta}^2_- \right) - \frac{N^2}{6a^2} U(\beta_+, \beta_-)  + \frac{1}{2} \left(\frac{1}{2}\dot{\phi}^2 - N^2 V(\phi)\right) = 0\,, \\
& \qquad \qquad \qquad \qquad \quad  \ddot{\beta}_+ + 3\frac{\dot{a}}{a}\dot{\beta}_+ - \frac{\dot{N}}{N}\dot{\beta}_+ + \frac{2}{3}\frac{N^2}{a^2} U_{,\beta_+} = 0\,, \label{betap} \\
& \qquad \qquad \qquad \qquad \quad  \ddot{\beta}_- + 3\frac{\dot{a}}{a}\dot{\beta}_- - \frac{\dot{N}}{N}\dot{\beta}_- + \frac{2}{3}\frac{N^2}{a^2} U_{,\beta_-} = 0\,. \label{betam}
\end{align}
Finally we have the equation for the scalar field,
\begin{align} \label{eomphi}
\ddot{\phi} + 3\frac{\dot{a}}{a}\dot{\phi} - \frac{\dot{N}}{N}\dot{\phi}+ N^2 V_{,\phi} = 0 \,.
\end{align}
One can simplify the equation for $a$ by plugging in the Friedman constraint \eqref{Friedman} into it. Then we get
\begin{align} \label{simplifieda}
\frac{\ddot{a}}{a} + \frac{1}{2}\left( \dot{\beta}^2_+ + \dot{\beta}^2_- \right) + \frac{1}{3} \left(\dot{\phi}^2 - N^2 V(\phi)\right) = 0\,.
\end{align}
Similarly, once we have a solution to the equations of motion, we can simplify the calculation of the value of the on-shell action by plugging in the Friedman equation \eqref{Friedman},
\begin{align} \label{onshell}
S_{on-shell} = - 4\pi^2 \int dt Na \left[ U(\beta_+, \beta_-) + a^2 V(\phi)\right]\,.
\end{align} 
In the numerical calculations, it turns out to be computationally favourable if one eliminates the $U(\beta_+,\beta_-)$ potential from the action. In that case the action becomes
\begin{align}
S_{on-shell} = 6\pi^2 \int dt \frac{a^3}{N} \left( -2 \dot{a}^2 + \frac{1}{2} (\dot{\beta}_+^2 + \dot{\beta}_-^2)  + \frac{1}{3} \dot{\phi}^2 \right)\,.
\end{align}

The potential for the anisotropy parameters $\beta_\pm$ is shown in Fig. \ref{fig:U}. For small $\beta$s it is given approximately by
\begin{align}
U(\beta_+,\beta_-) \approx -3 + 6\left( \beta_+^2 + \beta_-^2 \right),
\end{align}
and hence near the origin it has a circular symmetry. For larger anisotropies, the potential becomes exponentially steep and has the symmetry of an equilateral triangle, with one axis of symmetry being the $\beta_-=0$ line \cite{Misner:1969hg}. This forms the basis for describing the dynamics close to a cosmological singularity as that of a ball on this (or a closely related) effective triangular billiard table \cite{Damour:2002et} (with different boundary conditions, this system has also been quantised \cite{Kleinschmidt:2009cv,Koehn:2011ph}). Here we will however not need the billiards description.

\begin{figure}[h] 
\begin{center}
\includegraphics[width=0.6\textwidth]{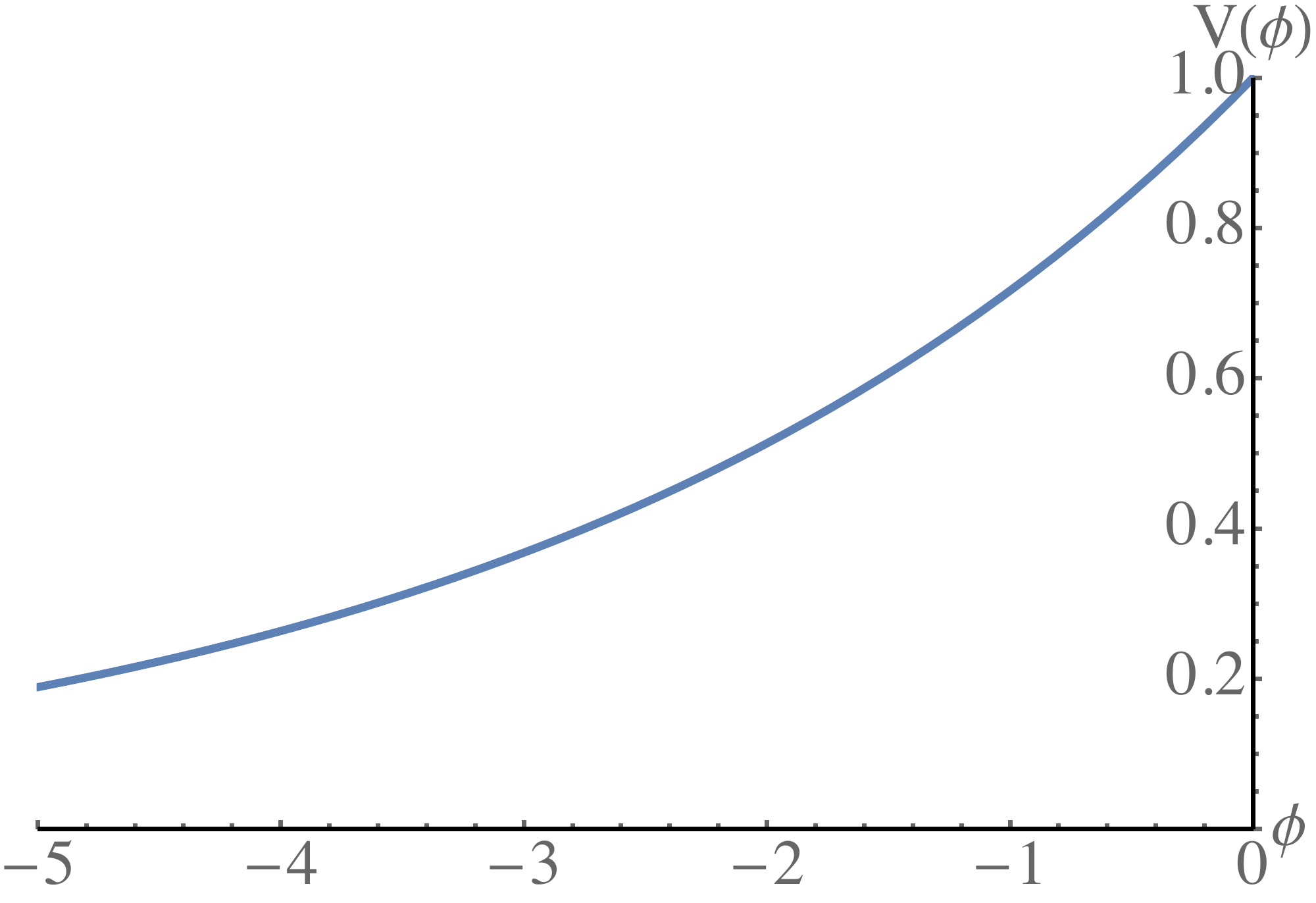}
\caption{The scalar field potential $V(\phi) = e^{c\phi}$. For our numerical examples, we chose $c=1/3$ and correspondingly $\epsilon = 1/18.$}
\label{fig:Pot}
\end{center}
\end{figure}

We will consider the scalar field potential to be of exponential form,
\begin{align}
V(\phi) = V_0 e^{c \phi},
\end{align}
as shown in Fig. \ref{fig:Pot} and with $c$ taken to be a positive constant. We will set $V_0=1,$ which can be achieved by shifting the origin of $\phi.$ The reason for choosing an exponential is that for such potentials the slow-roll parameter $\epsilon = c^2/2$ is constant (though it need not be small -- only the condition $\epsilon < 1$ is required for inflation to take place). Furthermore, with this potential, the theory has a classical scaling/shift symmetry. Indeed, if one performs the following transformations, with $\Delta \phi$ constant,
\begin{equation}
\phi  \equiv  \bar{ \phi} + \Delta \phi \;, \quad
g_{\mu\nu}  \equiv  e^{-c \Delta \phi} \bar{ g}_{\mu\nu} \;,\label{eq:metricscaling}
\end{equation}
one finds that the action changes only by an overall constant
\begin{equation} \label{eq:actionrescaled}
S = e^{-c\Delta \phi} \int d ^4x \sqrt{- \bar{ g}} \left( \frac{ \bar{R}}{2} - \frac{1}{2} \bar{g} ^{\mu \nu} \partial _{\mu} \bar{ \phi} \partial _{\nu} \bar{ \phi} - e^{c\bar\phi}\right) \;.
\end{equation}
This symmetry of the equations of motion is of great value in obtaining analytic approximations.

\section{Quantum Cosmology, Classicality and No-Boundary Conditions} \label{section:qc}

In order to define classicality in quantum cosmology it is useful to rewrite the complete Lorentzian action in the form
\begin{align}
S = 6\pi^2\int dtN\left[ \frac{1}{2} G_{AB} \left( \frac{1}{N}\frac{dq^A}{dt}\right) \left( \frac{1}{N}\frac{dq^B}{dt} \right) - \mathcal{U}(q^A) \right]
\end{align}
with $q^A = (a,\phi,\beta_+,\beta_-)$ and 
\begin{align}
G_{AB} = \text{diag}\left(-2a, \frac{1}{3}a^3, \frac{1}{2} a^3, \frac{1}{2} a^3 \right).
\end{align}
Then the associated Hamiltonian is given by
\begin{align} 
\mathcal{H} = \frac{1}{2}G^{AB}p_A p_B + \mathcal{U}\,,
\end{align}
with the canonical momenta $p_a = -2a\dot{a}$, $p_\phi = \frac{1}{3}a^3\dot{\phi}$, $p_{\beta_+}= \frac{1}{2}a^3\dot{\beta}_+$, $p_{\beta_-}= \frac{1}{2}a^3\dot{\beta}_-$, and where the total effective potential is given by
\begin{align}
\mathcal{U}(q^A) = a\,U(\beta_+,\beta_-) + a^3V(\phi).
\end{align}
Quantising this theory canonically by letting $p_A \rightarrow -i\hbar \frac{\partial}{\partial q^A}$ gives the quantum version of the Hamiltonian constraint (which in the classical theory is simply the Friedman equation), the Wheeler-DeWitt (WdW) equation
\begin{align}
\hat{\mathcal{H}}\Psi = \left( -\frac{\hbar^2}{2}\nabla^2 + \mathcal{U}\right)\Psi = 0\,,
\end{align}
where $\nabla^2 \equiv G^{AB} \nabla_A \nabla_B$ and where we have chosen the factor ordering such that the equation is covariant on superspace \cite{Halliwell:1988wc}. Let us choose a semi-classical ansatz for the wavefunction
\begin{align}
\Psi = e^{(-A +i\tilde{S})/\hbar}
\end{align}
with $A(a,\phi,\beta_+,\beta_-)$ and $\tilde{S}(a,\phi,\beta_+,\beta_-)$ real functions. Plugging this ansatz into the WdW equation and expanding in powers of $\hbar$ gives to leading order:
\begin{align}
-\frac{1}{2}G^{AB}\left( - \frac{\partial A}{\partial q^A} + i\frac{\partial \tilde{S}}{\partial q^A} \right) \left(- \frac{\partial A}{\partial q^B} + i\frac{\partial \tilde{S}}{\partial q^B} \right) + \mathcal{U} = 0
\end{align}
Thus if 
\begin{align} \label{classicality}
\frac{\partial \tilde{S}}{\partial q^A} \gg \frac{\partial A}{\partial q^A},
\end{align} 
i.e. if the phase of the wavefunction varies much faster than its amplitude for all degrees of freedom, we obtain the Lorentzian Hamilton-Jacobi equation (which specifies the classical dynamics)
\begin{align}
\frac{1}{2}G^{AB}\frac{\partial \tilde{S}}{\partial q^A}\frac{\partial \tilde{S}}{\partial q^B} + \mathcal{U} = 0\,,
\end{align}
as long as we identify $\tilde{S}$ with the classical action. With this identification, we also obtain the classical relation between the momenta and the action, 
\begin{align}
p_{A} = \frac{\partial \tilde{S}}{\partial q^A}\,,
\end{align}
and the behaviour of the wavefunction can be said to be classical since it is strongly peaked around classical solutions to the equations of motion. A possible probabilistic interpretation  of the wavefunction has been described by Vilenkin \cite{Vilenkin:1988yd} and relies on the conserved Klein-Gordon current
\begin{align}
J_B = -\frac{i}{2} (\Psi^* \nabla_B \Psi - \Psi \nabla_B \Psi^* )\,.
\end{align}
Evaluating this current for the semi-classical form of the wavefunction yields $J_B = e^{-2A} \nabla_B \tilde{S}$ and consequently
\begin{align}
\nabla^B \left( e^{-2A} \nabla_B \tilde{S} \right) = 0\,.
\end{align}
Vilenkin's prescription then is to specify a spacelike hypersurface in field space, and define approximately conserved relative probabilities $e^{-2A} \, n^B \nabla_B \tilde{S}$ where $n^B$ is the unit normal to the surface. 

The WdW equation admits many solutions. In order to know which one to pick, we need a theory of initial conditions. First recall that the path integral construction of the wavefunction, 
\begin{align} \label{PathIntegral}
\Psi(b,\chi,b_+,b_-) = \int_{\cal C} {\cal D}N  {\cal D}a  {\cal D}\phi  {\cal D}\beta_+  {\cal D}\beta_- \, e^{\frac{i}{\hbar}\int \mathrm{d}t\left[ p_A \dot{q}^A - N \mathcal{H} \right]}\,,
\end{align}
is equivalent to canonical quantisation, in the sense that the wavefunction thus constructed automatically solves the WdW equation (see e.g. \cite{Halliwell:1988wc}). Here the arguments of the wavefunction correspond to the specified field values on the final hypersurface. If we denote the time coordinate at the final hypersurface by $\tau_f,$ then the arguments are
\begin{equation} \label{finalvalues}
a(\tau_f) = b, \quad \phi(\tau_f) = \chi, \quad \beta_+(\tau_f) = b_+, \quad \beta_-(\tau_f) = b_-\,.
\end{equation}
In the definition \eqref{PathIntegral} the no-boundary proposal then restricts the class ${\cal C}$ of metrics over which the path integral is performed to be the class of compact, regular metrics admitting regular field configurations and having no boundary other than the final boundary just described. This restriction selects particular solutions of the WdW equation -- this is the sense in which the no-boundary proposal is indeed a theory of initial conditions. Below we will evaluate the path integral in the saddle point approximation, i.e. we will look for finite action solutions of the classical equations of motion satisfying the required boundary conditions. As is well known \cite{Lyons:1992ua}, with the ``no-boundary'' boundary conditions, these solutions must in fact be complex, although of course at the final boundary all field vales in \eqref{finalvalues} are required to be real. Given one saddle point, one can obtain others rather trivially, by taking either the complex conjugate or the time reverse  (or both) of a particular saddle point geometry. Hartle and Hawking then have a proposal as to which of these  saddle points should be retained \cite{Hawking:1981gb,Hartle:1983ai,Hawking:1983hj,Hartle:2008ng}. A second well-known theory of initial conditions is Vilenkin's tunnelling proposal. In that theory, the universe is also envisaged to tunnel from ``nothing'', and the regular tunnelling geometries satisfy the same no-boundary regularity condition. The difference with the approach of Hartle and Hawking is that the tunnelling boundary conditions select a different saddle point to be retained \cite{Vilenkin:1982de,Vilenkin:1983xq,Vilenkin:1984wp,Vilenkin:1986cy}. Since the various saddle points in question can be trivially obtained from one another, we will not dwell on distinguishing the two proposals below - our focus is on obtaining and characterising the saddle point geometries in the first place, and on the classicality properties of the wavefunction, which are issues that apply equally to both theories of initial conditions.

The no-boundary condition demands regularity at the so-called South Pole of the solution (i.e. where the volume of the universe is zero). In our case this corresponds to $a = 0,$ which we can set to be at $t = 0$. From the Friedman equation it is clear that at the South Pole
\begin{align}
\dot{a}^2 = \frac{N^2}{3} U(\beta_+, \beta_-)
\end{align}
must be satisfied. For small anisotropies, Eq. \eqref{anisotropypotential} implies that $U<0,$ and thus we see that the Friedman constraint forces us to complexify the fields (we will shortly see that in fact we need to take $U(\beta_+(t=0),\beta_-(t=0))=U(0,0)=-3$). The $\phi$ equation \eqref{eomphi} enforces
\begin{align}
\dot{\phi} = 0\,.
\end{align}
The $\beta$ equations \eqref{betap},\eqref{betam} give $U_{,\beta_+} = U_{,\beta_-} = 0$ which correspond respectively to 
\begin{align}
2 e^{ 2 \beta_+ } - e^{-\beta_+ - \sqrt{3}\beta_-} - e^{-\beta_+ + \sqrt{3}\beta_-} + 2e^{ -4 \beta_+ } - e^{2\beta_+ - 2\sqrt{3}\beta_-} - e^{2\beta_+ + 2\sqrt{3}\beta_-} = 0\,, \\
- e^{-\beta_+ - \sqrt{3}\beta_-} + e^{-\beta_+ + \sqrt{3}\beta_-} -  e^{2\beta_+ - 2\sqrt{3}\beta_-} + e^{2\beta_+ + 2\sqrt{3}\beta_-} = 0\,.
\end{align}
These equations allow six complex solutions given by
\begin{align}
(e^{\beta_+},e^{\sqrt{3}\beta_-}) &= \left\{ (1,1),(-1,-1),(-(-1)^{1/3}, 1),((-1)^{1/3}, -1),((-1)^{2/3}, 1),(-(-1)^{2/3},-1)  \right \}\,.
\end{align}
It is instructive to analyse the form of the metric near the South Pole for these values. Inserting the values of the first two solutions yields
\begin{align} \label{SPa1}
ds^2_{SP} \approx -N^2 dt^2 + a^2\left(\sigma_1^2 + \sigma_2^2 + \sigma_3^2\right)\,.
\end{align}
Solutions 3 and 4 give 
\begin{align}
ds^2_{SP} &\approx -N^2 dt^2 - \left( \frac{1}{2} +i\frac{\sqrt{3}}{2} \right)a^2\left(\sigma_1^2 + \sigma_2^2 + \sigma_3^2\right)\,,
\end{align}
while the last pair of solutions give
\begin{align}
ds^2_{SP} &\approx -N^2 dt^2 + \left( -\frac{1}{2} +i\frac{\sqrt{3}}{2} \right)a^2\left(\sigma_1^2 + \sigma_2^2 + \sigma_3^2\right)\,.
\end{align}
Since we are allowing for complex scale factors $a$ when searching for instanton solutions, all these cases are in fact equivalent, and we may simply use \eqref{SPa1}.

Even though we defined the path integral in real/Lorentzian time, we just saw that the boundary conditions force us to consider complex solutions as saddle points of the action. This means that from here on we should consider the time variable to be complex. To make contact with the existing literature on no-boundary instantons, we will take our complexified time variable to be given by $\tau,$ such that $Im(\tau)=t.$ Thus one may think of the real part of $\tau$ as denoting the Euclidean time direction, and the imaginary part as real time. The regularity of the field equations near the South Pole then translates into the following series expansions up to ${\cal O}(\tau^5)$
\begin{align}
a &= \tau -\frac{1}{18} V_0 e^{c\phi_{SP}} \tau^3+\frac{1}{8640}((-216 ( \beta_{SP+}'' )^2-216 ( \beta_{SP-}'' )^2+(8-27 c^2) V_0^2 e^{2 c \phi_{SP}}) \tau^5 +\cdots \label{seriesa} \\
\phi &= \phi_{SP}+\frac{c}{8}  V_0 e^{c \phi_{SP}} \tau^2+\frac{c (2+3 c^2)}{576}  V_0^2 e^{2 c \phi_{SP}} \tau^4+\cdots \\
\beta_+ &= \frac{1}{2} \beta_{SP+}'' \tau^2 + \frac{1}{144} (45 ( \beta_{SP-}'' )^2+ \beta_{SP+}'' (-45\beta_{SP-}'' +7 V_0 e^{c \phi_{SP}})) \tau^4+\cdots \\
\beta_- &= \frac{1}{2}  \beta_{SP-}'' \tau^2 + \frac{1}{144}  \beta_{SP-}'' (90  \beta_{SP+}''+7 V_0 e^{c \phi_{SP}}) \tau^4 +\cdots \label{seriesbn}
\end{align}
These series expansions are needed to form a well-defined numerical problem. We can see that the instantons are characterised by the three complex numbers
\begin{equation} \label{SPvalues}
\phi_{SP}, \quad \beta_{SP+}'', \quad \beta_{SP-}''\,,
\end{equation}
representing the scalar field value, and the values of the second derivatives of the anisotropy functions, at the South Pole. The no-boundary condition forces the anisotropy functions and their first derivatives to be zero at the no-boundary point, but allow for a non-trivial second derivative. In this way anisotropies can develop.

\section{Results}

\subsection{Existence and basic features of anisotropic instantons}

\begin{figure}[h] 
\begin{center}
\includegraphics[width=0.75\textwidth]{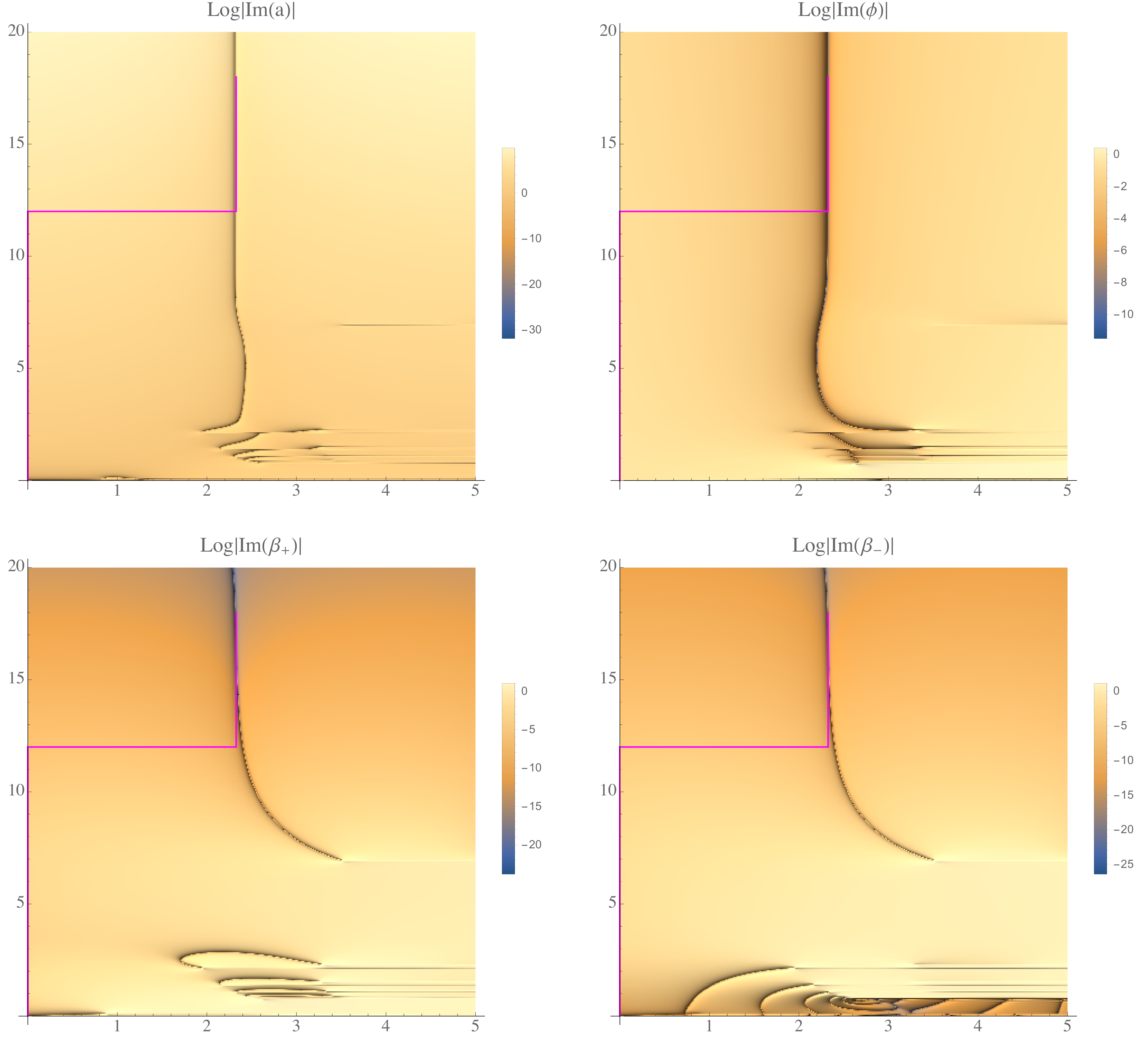}
\caption{An example of an anisotropic instanton, optimised to reach the real values $b = 10000, \chi= -2, b_+=1, b_-=1$ on the final boundary. These values are reached at $\tau_f = 2.32705 + 17.9932\,i$, with the South Pole values $\phi_{SP}=0.942081 - 0.554398\,i, \beta_{SP+}'' = -0.926417 + 0.173177\, i, \beta_{SP-}'' = -0.00373004 + 0.000697265\, i.$ We have drawn an example of a ``good'' contour of integration in magenta, which avoids the singularities and their associated branch cuts visible in the lower right part of the figures. For a detailed description of the figure, see the main text.}
\label{fig:afterdip}
\end{center}
\end{figure}

\begin{figure}[h] 
\begin{center}
\includegraphics[width=0.65\textwidth]{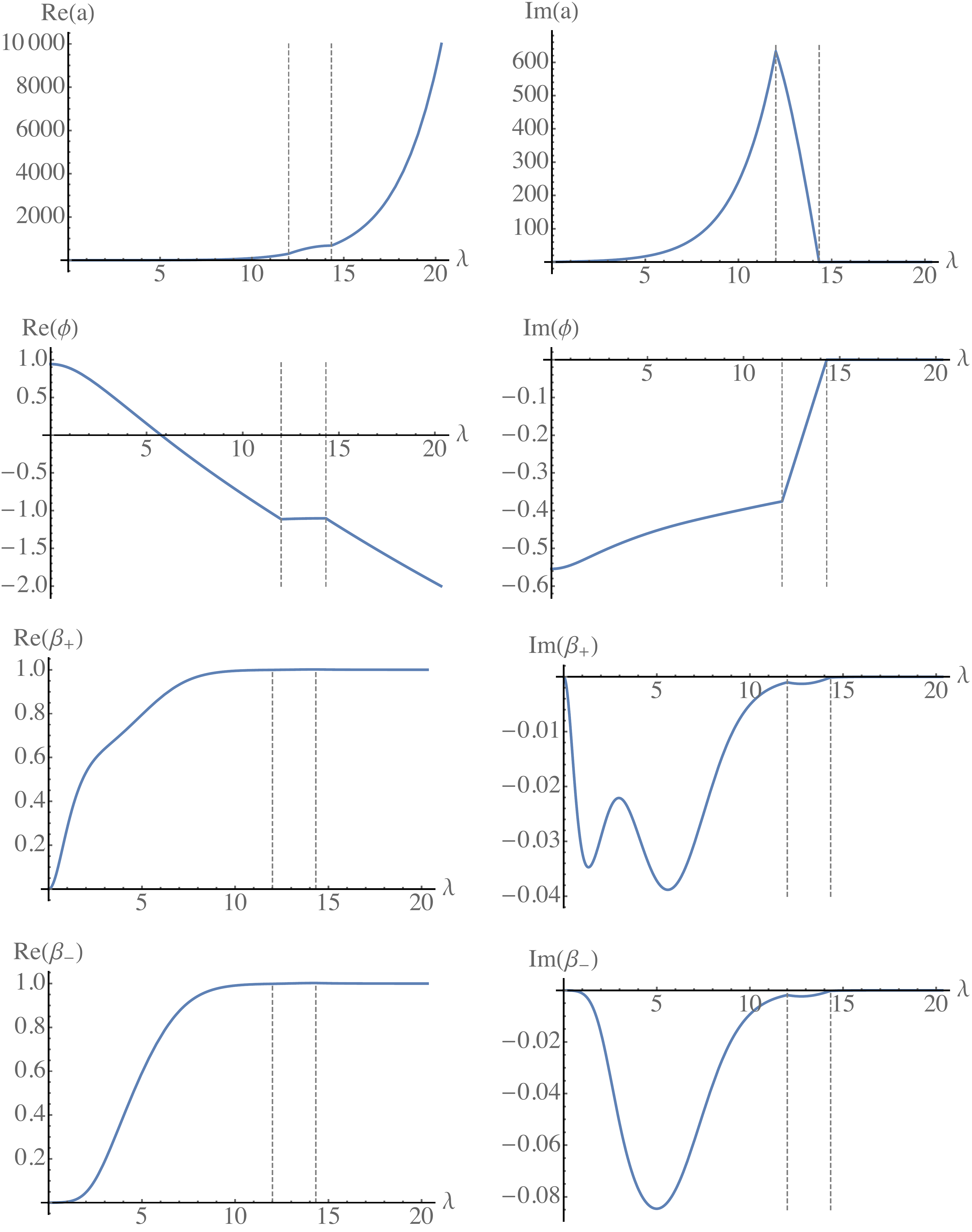}
\caption{The evolution of the fields $a, \phi, \beta_+, \beta_-$ along the contour shown in magenta in Fig. \ref{fig:afterdip}. The contour has been parameterised with a monotonically increasing parameter $\lambda,$ and the dashed lines indicate the locations where the contour changes direction. Note that the fields approach real values on the final, vertical part of the contour. The inflationary attractor ensures that this is possible simultaneously for all fields. Also note that the anisotropy functions $\beta_\pm$ start out at zero, as they must to satisfy the no-boundary conditions, then grow to complex values and and eventually settle at the desired real values.}
\label{fig:fieldevolution}
\end{center}
\end{figure}

We can now look for solutions satisfying the no-boundary conditions \eqref{seriesa} - \eqref{seriesbn} while approaching the desired real values of $b,\chi, b_+, b_-$ on the final hypersurface at some $\tau_f.$ In order to find such solutions we have the freedom of adjusting the contour and the South Pole values \eqref{SPvalues}. We find these values by implementing a numerical Newtonian optimisation algorithm. An example of an anisotropic instanton, optimised to reach the values $(b=10000,\chi=-2,b_+=1,b_-=1)$ on the final boundary is shown in Fig. \ref{fig:afterdip}. What we show in the figure are relief plots of the imaginary parts of the functions $a(\tau), \phi(\tau), \beta_\pm(\tau)$ over the complex time plane $\tau$ with $\tau=0$ corresponding to the South Pole where the no-boundary conditions are implemented. More precisely, we are plotting the logarithm of the absolute value of the imaginary part of these functions, such that small imaginary part corresponds to very negative values and thus very dark points. The dark lines thus represent the locus where the fields are essentially real. These plots are obtained by solving the equations of motion, starting from the South Pole, going upwards along the imaginary $\tau$ axis to a fixed height first, and then branching out horizontally to a dense series of points on a horizontal line. Then this procedure is repeated for a slightly higher horizontal line, until a dense grid of points is obtained, covering the desired region of the complex time plane.

\begin{figure}[h] 
\begin{center}
\includegraphics[width=0.5\textwidth]{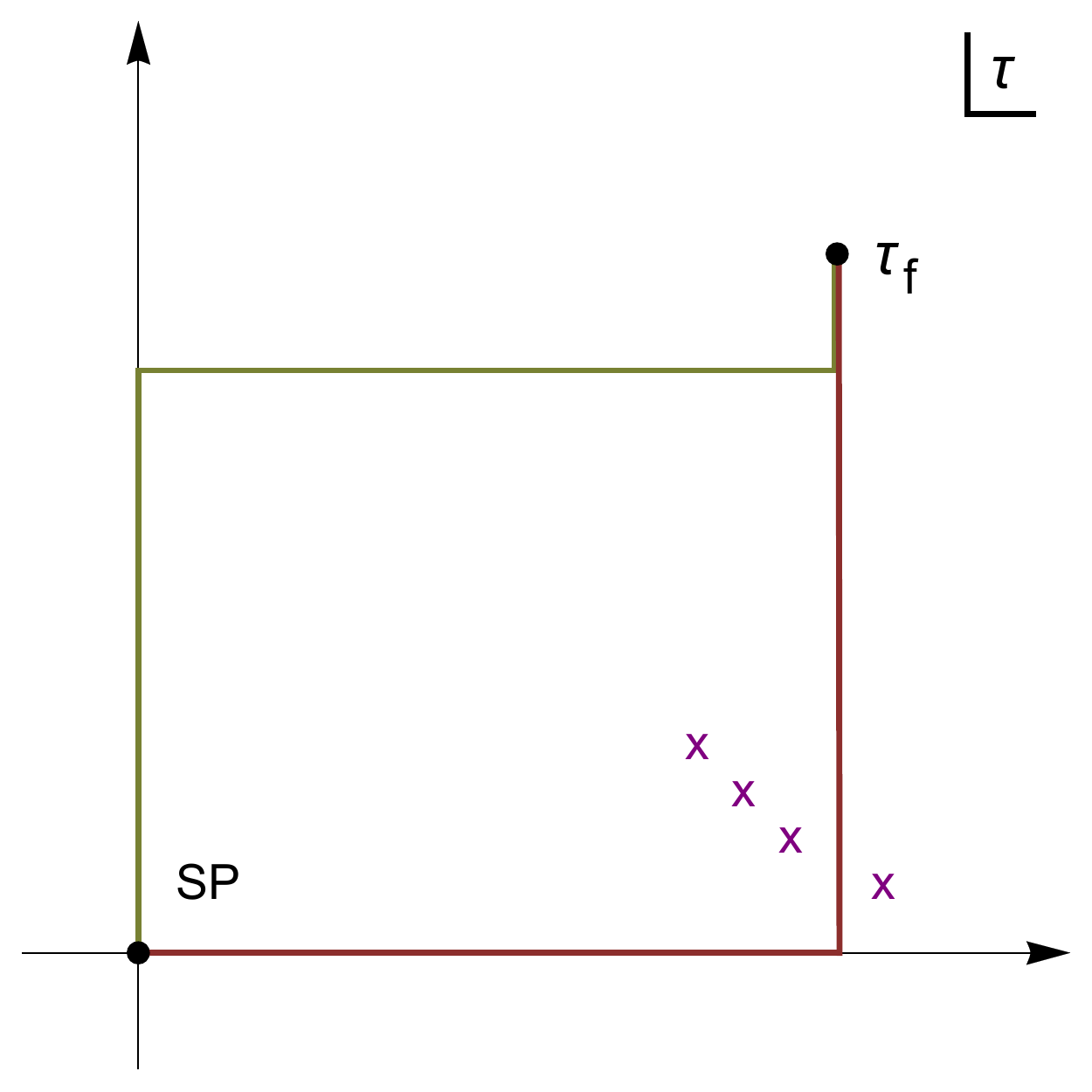}
\caption{Due to the presence of singularities (marked by purple crosses), we cannot choose the standard ``Hawking'' contour in the complex time plane (in red), as this contour would not have yielded a solution with the desired boundary conditions. Instead we have to use a modified contour such as the one in green. Shown here is the complexified Euclidean time plane $\tau,$ with the South Pole at $\tau=0$ and the final boundary conditions imposed at $\tau_f.$}
\label{fig:contour}
\end{center}
\end{figure}

Our procedure thus implicitly entails a choice of contour along which the equations of motion are solved. This contour is different from the type of contour usually employed for no-boundary instantons, see Fig. \ref{fig:contour}. The usual contour runs out horizontally along the real $\tau$ axis (along which the solution is approximately that of a Euclidean sphere) and then up, parallel to the imaginary $\tau$ axis, to the final location $\tau_f$ where the desired field values are reached. In the isotropic case,  the solution along this last contour then corresponds to a portion of de Sitter space. However, when significant anisotropies are included, this standard type of contour is no longer viable, as singularities develop and the standard contour would in fact take us to a different sheet of the solution function. Along this new sheet we have checked and found that the final real values $(b,\chi,b_\pm)$ are not reached. But we can avoid the singularities by running the contour first up along the imaginary $\tau$ axis, and then horizontally across. An example of such a ``good'' contour is shown by the magenta line in Fig. \ref{fig:afterdip}, and the evolution of the fields along this contour is shown in Fig. \ref{fig:fieldevolution}.

Note that the presence of additional singularities is not really surprising: the anisotropies lead to an increased energy density, which favours a decelerating scale factor (see Eq. \eqref{simplifieda}) and thus favours gravitational collapse. In regions where the scale factor $a$ shrinks, a singularity can only be avoided if the homogeneous curvature dominates over the anisotropies, since the homogeneous curvature can induce a bounce analogous to that present in the closed slicing of de Sitter space. This however will generically not occur, as the energy density of the homogeneous curvature scales as $1/a^2$ while that of the anisotropies scales as $1/a^6.$ Thus we may generically expect singularities to form in regions where the scale factor shrinks, and consequently it is only natural that we see many additional singularities in the anisotropic case. (See also \cite{Bramberger:2017cgf} for tunnelling solutions which circumvent singularities in a similar manner.)

\begin{figure}[h] 
\begin{minipage}{0.5\textwidth}
		\includegraphics[width=0.9\textwidth]{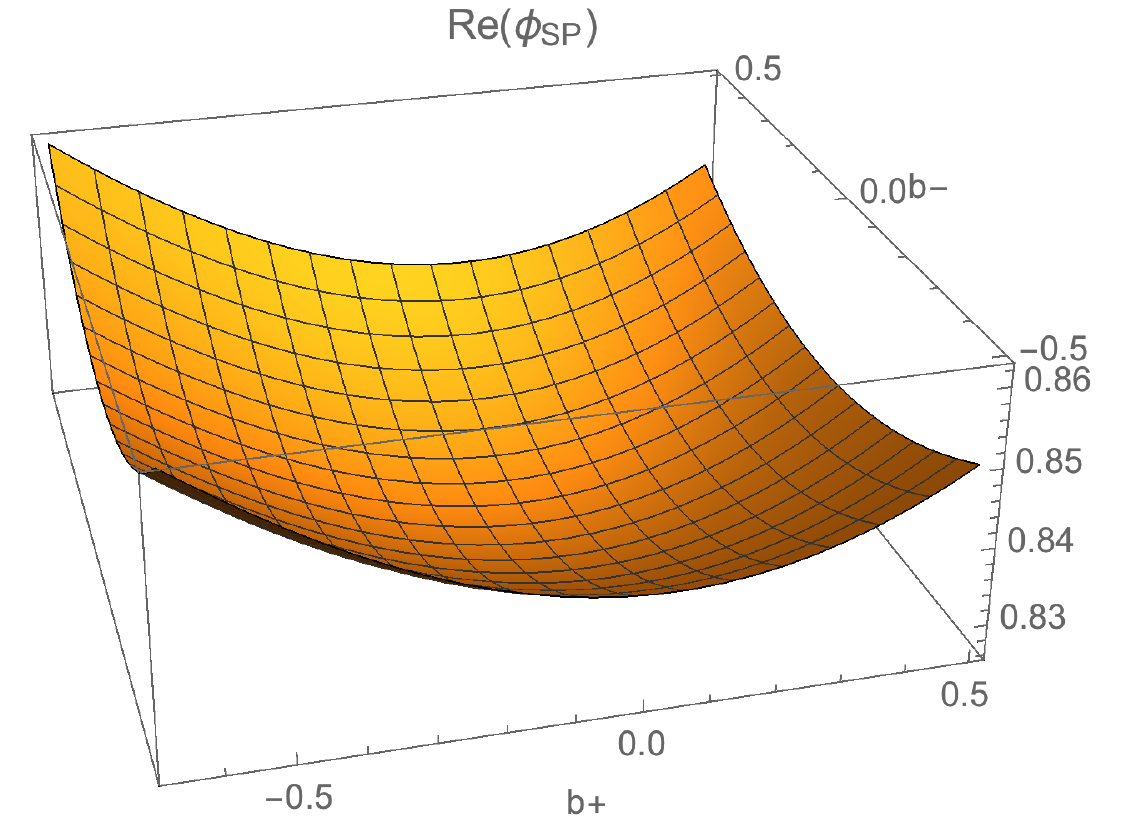}
	\end{minipage}%
	\begin{minipage}{0.5\textwidth}
		\includegraphics[width=0.9\textwidth]{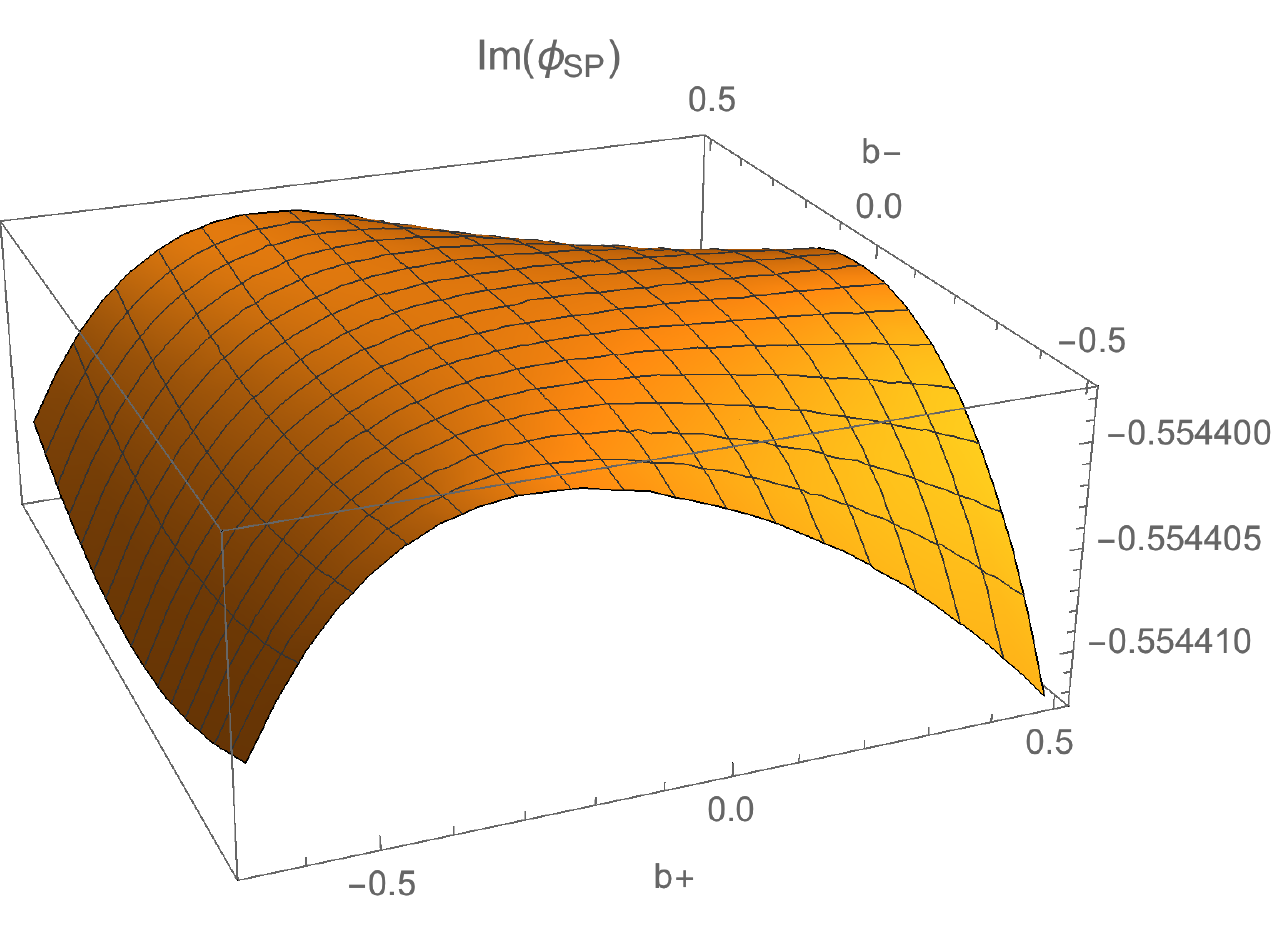}
	\end{minipage}%
	\caption{Real and imaginary parts of $\phi_{SP}$ at the South Pole, in terms of the final, real values of $b_\pm$ indicated, and for $b=100,\, \chi=-1/2$. Note that $Im(\phi_{SP})$ varies only over a very small range.}
	\label{fig:phiSP}
\end{figure}

\begin{figure}[h] 
\begin{minipage}{0.5\textwidth}
		\includegraphics[width=0.9\textwidth]{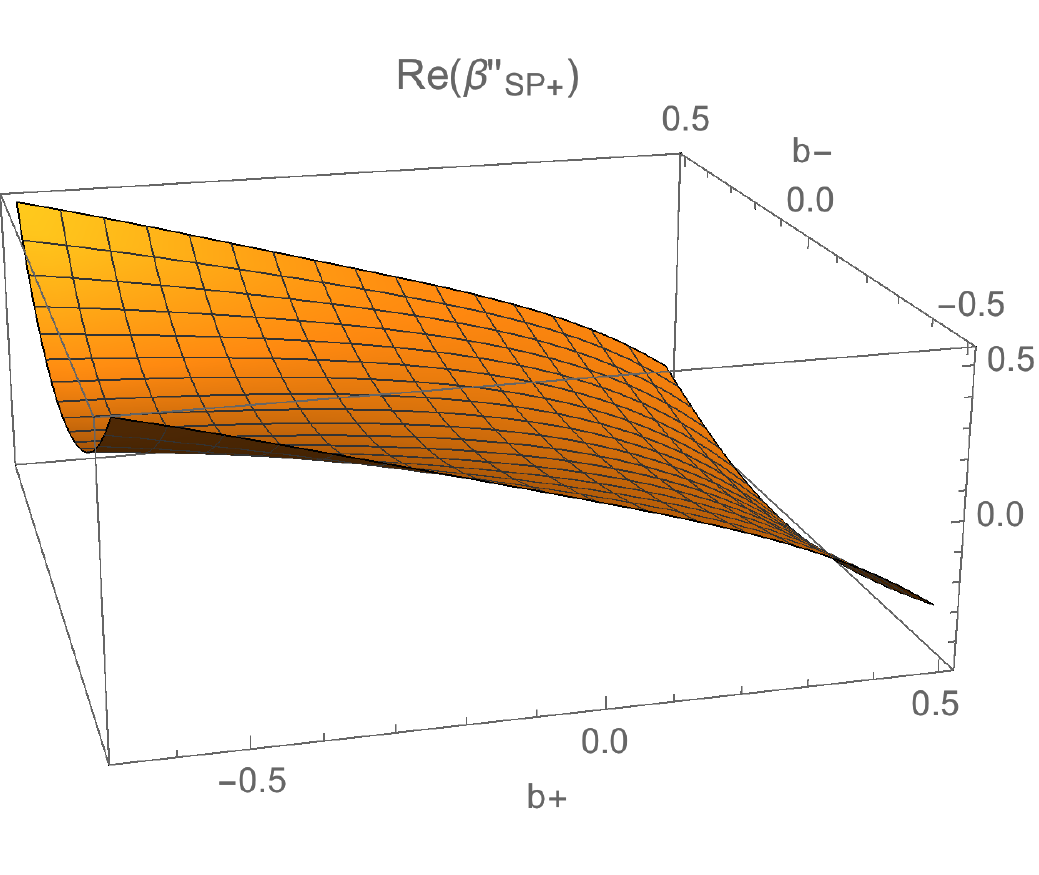}
	\end{minipage}%
	\begin{minipage}{0.5\textwidth}
		\includegraphics[width=0.9\textwidth]{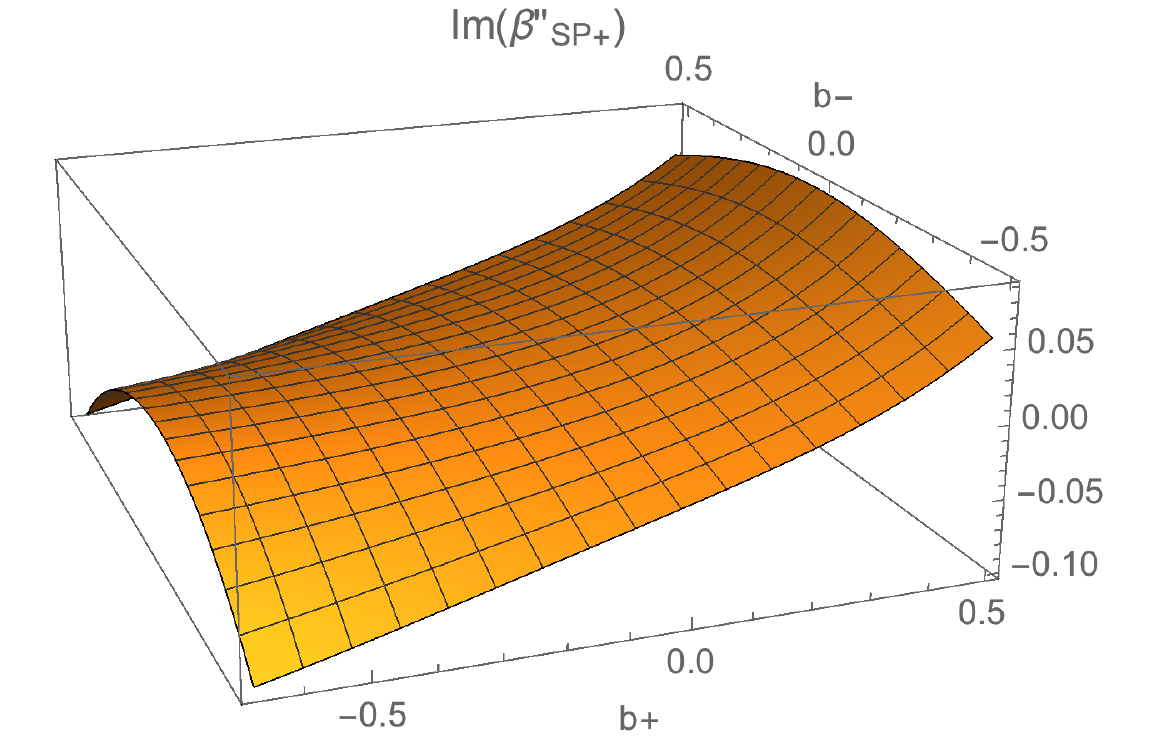}
	\end{minipage}%
	\caption{Real and imaginary parts of $\beta_{SP+}''$ at the South Pole, in terms of the final, real values of $b_\pm$ indicated, and for $b=100,\, \chi=-1/2$.}
	\label{fig:bpSP}
\end{figure}

\begin{figure}[h] 
\begin{minipage}{0.5\textwidth}
		\includegraphics[width=0.9\textwidth]{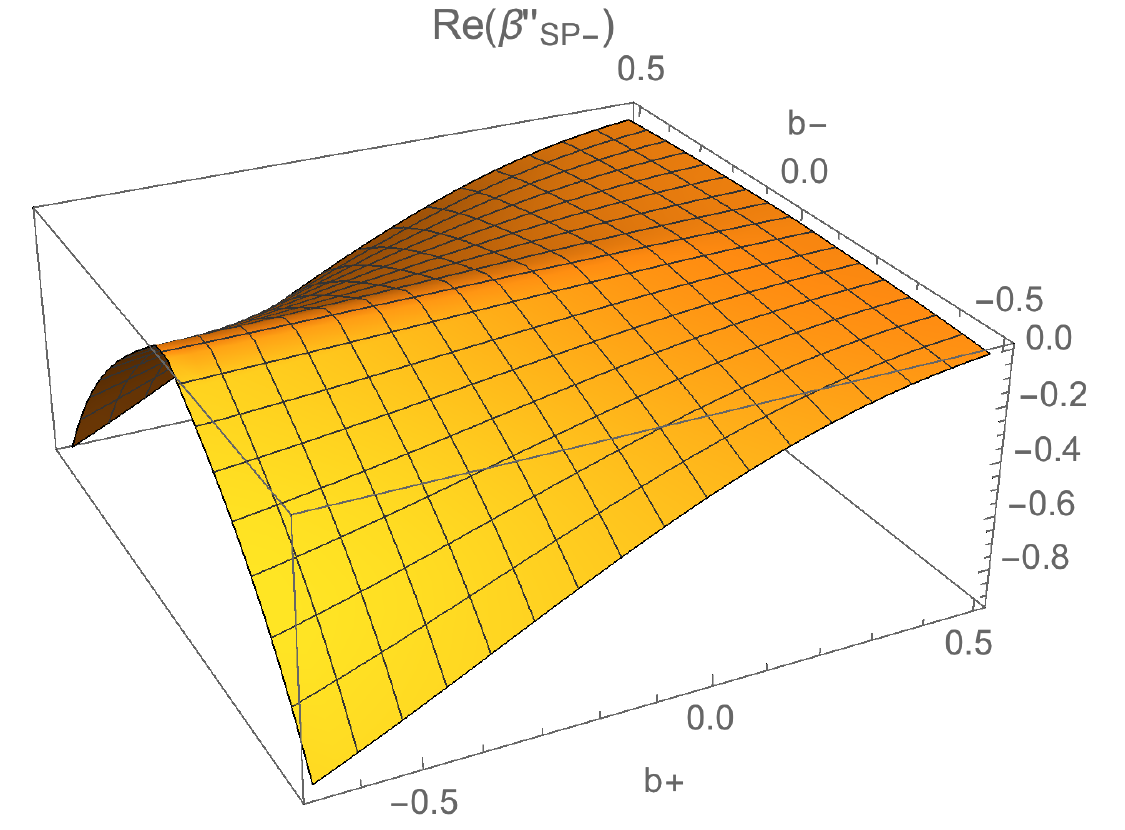}
	\end{minipage}%
	\begin{minipage}{0.5\textwidth}
		\includegraphics[width=0.9\textwidth]{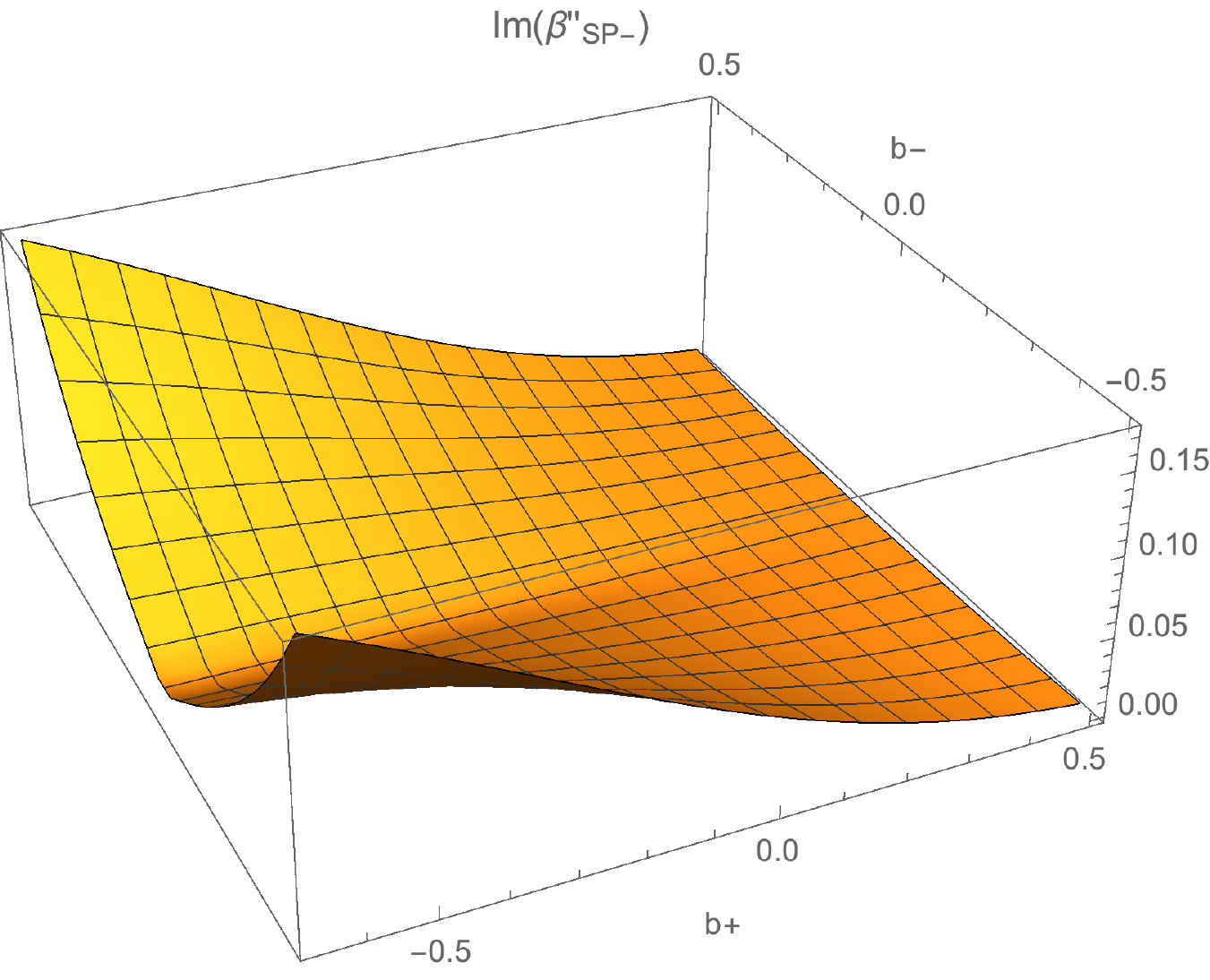}
	\end{minipage}%
	\caption{Real and imaginary parts of $\beta_{SP-}''$ at the South Pole, in terms of the final, real values of $b_\pm$ indicated, and for $b=100,\, \chi=-1/2$.}
	\label{fig:bnSP}
\end{figure}

\begin{figure}[h] 
\begin{minipage}{0.5\textwidth}
		\includegraphics[width=0.9\textwidth]{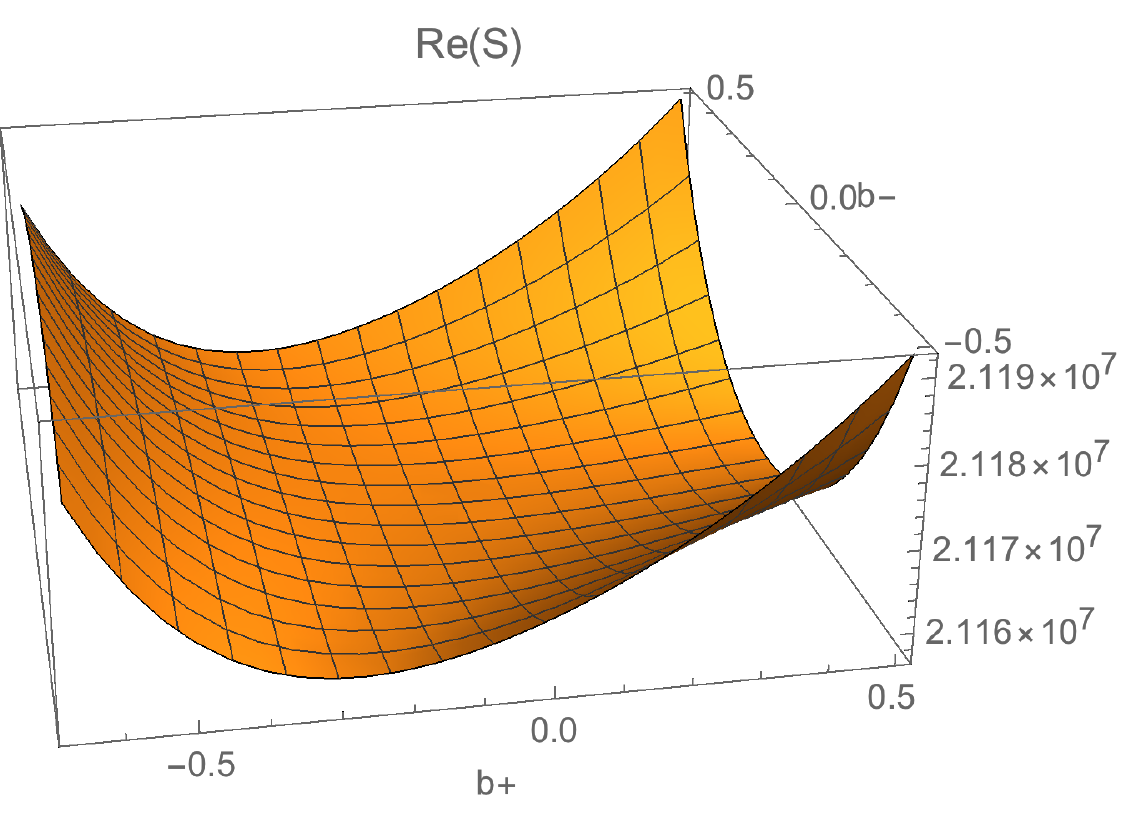}
	\end{minipage}%
	\begin{minipage}{0.5\textwidth}
		\includegraphics[width=0.9\textwidth]{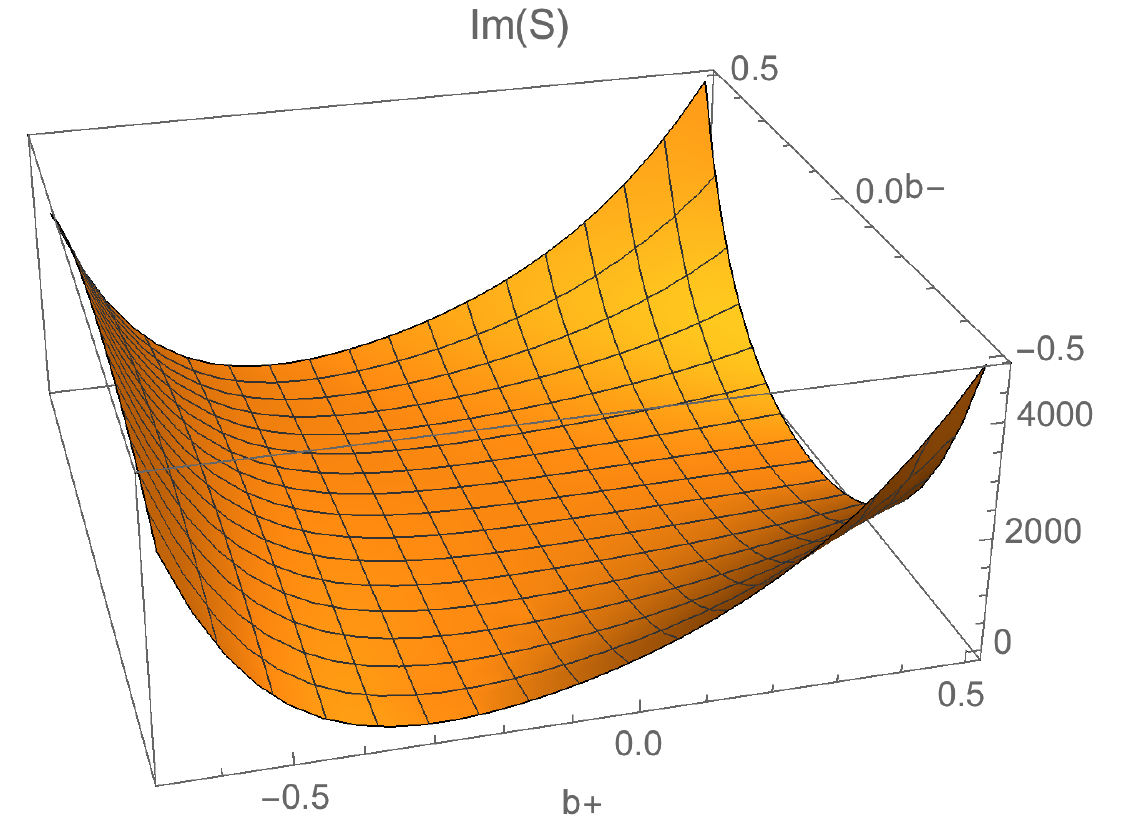}
	\end{minipage}%
	\caption{Real and imaginary parts of the action $S$, in terms of the final, real values of $b_\pm$ indicated, and for $b=100,\, \chi=-1/2$.}
	\label{fig:action}
\end{figure}

With the right contour, we can now construct anisotropic instantons over large ranges of values, where we only seem to be limited in the range by the computational time it takes to optimise the instantons. As an example, we show the South Pole values and values of the action for instantons optimised to reach $(b=100, \chi= -1/2),$ with the anisotropy parameters ranging from $-7/10 \leq b_+ \leq +1/2$ and $-1/2 \leq b_- \leq + 1/2.$ These ranges coincide with the ranges for the potential $U(\beta_+,\beta_-)$ shown in Fig. \ref{fig:U}. The optimised South Pole values $\phi_{SP}, \beta_{SP+}'', \beta_{SP-}''$ are shown in Figs. \ref{fig:phiSP} - \ref{fig:bnSP}, while the action is shown in Fig. \ref{fig:action}. The figures clearly reflect the expected $b_- \rightarrow - b_-$ symmetry that comes with this choice of coordinates. Note that the South Pole values of the scalar field vary little as the anisotropies are increased, and in particular the imaginary part stays essentially constant. Note also that for a pure $b_+$ deformation the values of $\beta_{SP-}''$ stay close to zero, and to a somewhat lesser extent this is also true for the $\beta_{SP+}''$ values when considering pure $b_-$ deformations. This indicates that there is not much ``rotation'' (or mixing between $\beta_+$ and $\beta_-$) of the instantons between the South Pole and the final hypersurface. Regarding the action in Fig. \ref{fig:action}, we can see that the real part of the action is very large, which is as expected since a classical history has been reached. The imaginary part of the action, which can be thought of as the ``quantum'' part, is much smaller but increases steeply for larger anisotropies. We note also that it changes sign: the minimum is located at $b_+=b_-=0$ where $Im(S)=-79.769844,$ while for large anisotropies the imaginary part of the action becomes large and positive. We will further comment on this feature in the discussion section.

Having constructed anisotropic instantons over a significant range of anisotropy parameters, we should note an interesting consequence of the shift/scaling symmetry \eqref{eq:metricscaling}. Given a solution such as the ones we have just described, a shifted instanton with final values
\begin{align} \label{shift2}
b \rightarrow b \, e^{-\frac{c}{2}\Delta\chi}, \quad \chi \rightarrow \chi + \Delta \chi, \quad b_+ \rightarrow b_+, \quad b_- \rightarrow b_-\,,
\end{align}
can be obtained from the following South Pole values 
\begin{align}
\phi_{SP} &\rightarrow \phi_{SP}+ \Delta\chi \\
\beta_{SP+}'' &\rightarrow \beta_{SP+}'' \, e^{c\Delta\chi} \\
\beta_{SP-}'' &\rightarrow \beta_{SP-}'' \, e^{c\Delta\chi} \\
\tau_f &\rightarrow \tau_f \, e^{-\frac{c}{2}\Delta\chi}\,,
\end{align}
where we also included the shifted time coordinate of the final hypersurface. Here $\Delta\chi$ is an arbitrary real number, und thus a one-parameter family of instantons with the same anisotropy parameters, but different scale factor and scalar field values can be obtained. These shifted instantons belong to different classical histories. Interestingly, starting from a specific instanton, one can use these relations to construct an instanton with the same anisotropies but with a much larger value of the scale factor. But evolving that new history back in time to the original scale factor one realises that this has shifted to a history with much larger anisotropies (as measured at a reference scale factor value). Given that using the formulae above we can shift the scale factor by an arbitrary amount, this means that we can obtain histories with an arbitrarily large anisotropies, along a one-parameter set of deformations. Together with our grids in Figs. \ref{fig:phiSP}, \ref{fig:bpSP} and \ref{fig:bnSP} this strongly suggests that, at least in the case of a constant equation of state, there is no limit to how large the anisotropies can be. 

The visual methods employed here, and which were developed in \cite{Battarra:2014kga,Battarra:2014xoa,Bramberger:2016yog}, have the great advantage of allowing one to see by eye where the singularities are located, and thus clearly show in what manner the choice of contour is crucial. With the right contour, we have been able to construct anisotropic instantons with any desired final anisotropy parameters. Thus we suspect that the obstruction to constructing instantons with large anisotropies reported in \cite{Fujio:2009my} might have been due to the fact that the authors used the standard contour, and thus inadvertently landed on a wrong sheet of the solution function.

\subsection{Scaling of the classicality conditions}

\begin{figure}[h] 
\begin{center}
\includegraphics[width=0.95\textwidth]{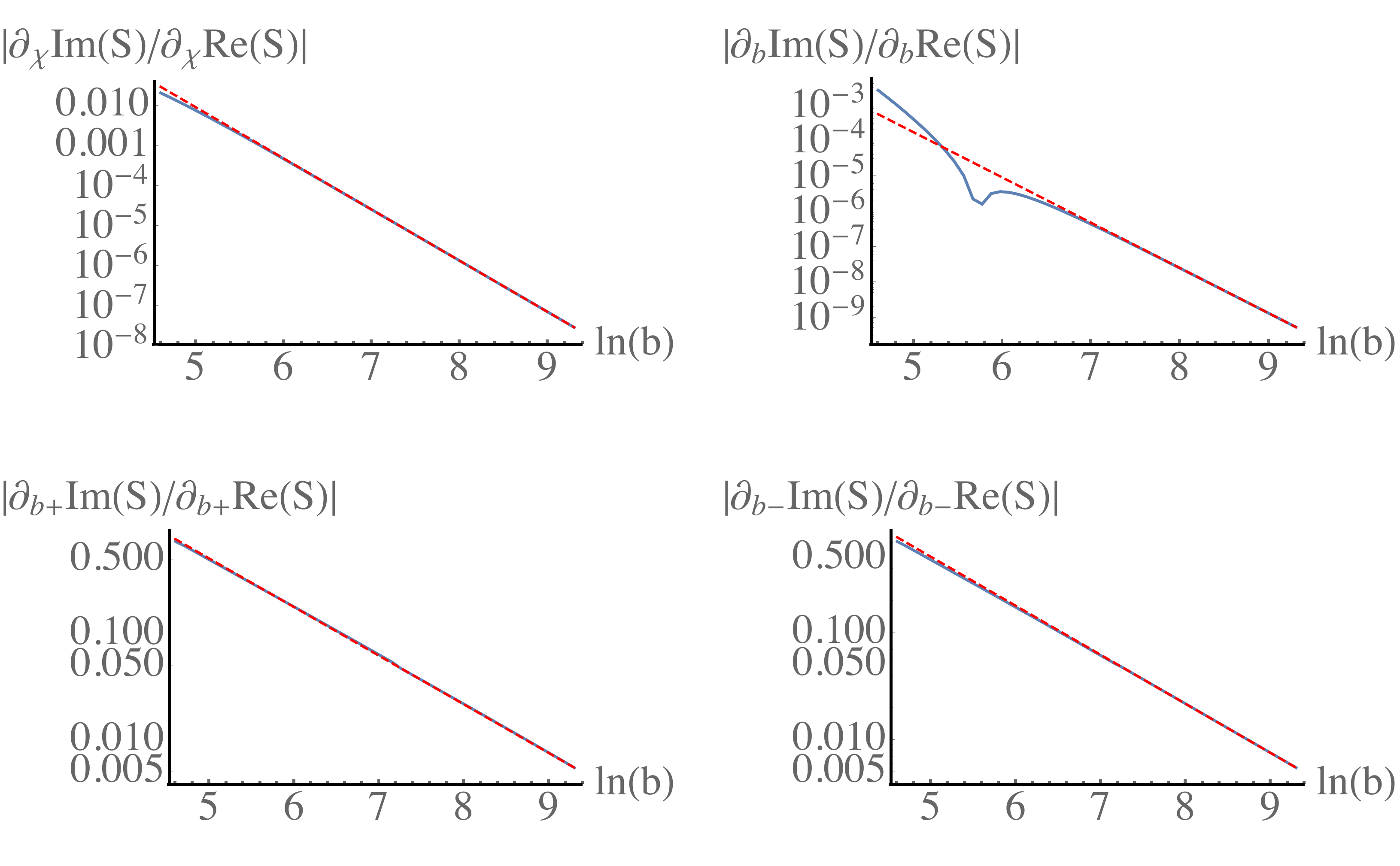}
\caption{Plots of the WKB classicality conditions (in blue) and their asymptotic scaling behaviour (red dashed lines). For the classicality conditions involving the scale factor $b$ and the scalar field $\chi$ the red dashed lines are proportional to $b^{-3+\epsilon},$ while for the relations involving the anisotropy functions $b_\pm$ the fitted red dashed lines are proportional to $b^{-1-\epsilon}.$ Thus the anisotropies cause the wavefunction to become classical more slowly than in the isotropic case.}
\label{fig:wkb}
\end{center}
\end{figure}

So far, we have discussed the instanton solutions that are required to approximate the wavefunction \eqref{PathIntegral} in the saddle point approximation. But it is important to realise that the instantons themselves do not represent the physical spacetime (which is also why it is unproblematic that they are complex valued) -- rather all the physics must be deduced from the wavefunction itself. The most basic question we can ask is whether the wavefunction thus calculated predicts a classical spacetime. We can analyse this question using the WKB classicality conditions reviewed around Eq. \eqref{classicality}. To evaluate whether the amplitude of the wavefunction evolves slowly compared to its phase, we must first find out how the action changes as the boundary conditions $(b, \chi, b_\pm)$ of the wavefunction are varied, i.e. we must evaluate the wavefunction along a classical history. Moreover, to evaluate the partial derivatives w.r.t the fields we must also evaluate the wavefunction with small changes in the individual fields, so that we may approximate the derivatives by finite differences. Thus we must evaluate $\Psi[b(\lambda), \chi(\lambda), b_+(\lambda), b_-(\lambda)]$ for a sequence of time steps, where $[b(\lambda), \chi(\lambda), b_+(\lambda), b_-(\lambda)]$ denotes a classical history parameterised by a time coordinate $\lambda,$ and also the slightly shifted instantons $\Psi[b+\delta b, \chi, b_+, b_-],$ $\Psi[b, \chi + \delta \chi, b_+, b_-],$ $\Psi[b, \chi, b_+ + \delta b_+, b_-]$ and $\Psi[b, \chi, b_+, b_- + \delta b_-]$ at each time step. Then we can form the WKB conditions
\begin{align}
{\cal{WKB}}_{\, q^A} \equiv \frac{\partial_A Im(S)}{\partial_A Re(S)}\,, \quad q^A=(b, \chi, b_+, b_-)\,,
\end{align}
which are shown in Fig. \ref{fig:wkb}. The numerical results for the WKB conditions are given by the blue lines, while the red dashed lines indicate fitting functions. There are a few points to note: the most obvious feature is that the WKB conditions become better and better satisfied as the universe expands. Thus the wavefunction really does predict a classical spacetime at large values of the scale factor. The seond point to note is that the WKB conditions approach a scaling law, since the log-log plots approach straight lines. Interestingly however, the four conditions do not all approach the same scaling. The conditions involving derivatives of the scale factor $b$ and the scalar field $\chi$ approach the scaling relation
\begin{align} \label{wkbbchi}
{\cal{WKB}}_{\, b,\chi} & \propto \frac{1}{b^{3-\epsilon}} \propto e^{-\frac{3-\epsilon}{1-\epsilon}{\mathcal N}}\,, 
\end{align}
which is the same scaling that one obtains for isotropic inflationary universes (here ${\mathcal N}$ denotes the number of e-folds of evolution, $\mathrm{d} {\mathcal N} \equiv \mathrm{d} \ln (aH)$). This is perhaps not so surprising, since the anisotropies are diluted away at late values of the scale factor. 
For a small slow-roll parameter $\epsilon,$ one has ${\cal{WKB}}_{\,b,\chi} \sim b^{-3},$ i.e. the classicality conditions are satisfied in inverse proportion to the volume generated by inflation. These relations were proven analytically for the isotropic case in \cite{Lehners:2015sia}. 

For the WKB conditions involving derivatives of the anisotropy functions $b_\pm,$ we obtain a different scaling law, namely
\begin{align}
{\cal{WKB}}_{\,b_+,b_-} & \propto \frac{1}{b^{1+\epsilon}} \propto e^{-\frac{1+\epsilon}{1-\epsilon}{\mathcal N}}\,.
\end{align} 
This is a substantially slower fall-off than that in Eq. \eqref{wkbbchi}, and for small slow-roll parameter $\epsilon$ one approximately finds ${\cal{WKB}}_{\,b_\pm} \sim b^{-1},$ that is to say the classicality conditions only become satisfied in inverse proportion to the linear size of the universe. Thus the anisotropies slow down the approach to classicality. 

We can derive this asymptotic scaling analytically. For this we need to derive the behaviour of the fields at large scale factor. At late times, the inflationary attractor is reached, and the energy density in the anisotropies is diluted as $1/a^6.$ Thus the anisotropies will only act as a small perturbation. Because of the attractor, successive constant time slices of a single instanton will correspond with great accuracy to a series of subsequent instantons for a wavefunction evaluated on the corresponding classical history. In an exponential scalar potential the scale factor will approach the inflationary attractor solution
\begin{align}
b &= b_0 t^{1/\epsilon} \,,
\end{align}
where $t$ is the Lorentzian time coordinate and $b_0$ is a constant. At large scale factor, we can consider the anisotropy equations of motion at linear order in the anisotropy functions,
\begin{align}
\ddot{b}_\pm + \frac{3}{\epsilon t}\dot{b}_\pm + \frac{8}{b_0^2 t^{2/\epsilon}}b_\pm = 0\,.
\end{align}
These equations can be solved asymptotically in a series expansion, giving
\begin{align} \label{AsymptoticBetas}
b_\pm(t) = b_{\infty \pm} \left(1 + \frac{4\epsilon^2}{b_0^2 (1-\epsilon^2)} t^{2-\frac{2}{\epsilon}} + \cdots \right)\,.
\end{align}
Here $b_{\infty \pm}$ are the asymptotic values of the anisotropy parameters reached at $t \rightarrow \infty.$

It is interesting to see how this solution transforms under the shift-scaling symmetry \eqref{shift2} that arises for exponential potentials. This symmetry only affects the time coordinate $t$ and the scale factor $b$ in the metric, and not the anisotropy parameters $b_\pm,$ so that we have
\begin{align}
\bar{b}_\pm (\bar{t}) &= b_\pm (e^{-\frac{c}{2}\Delta \chi}t) \\ &= b_{\infty \pm} \left(1 + \frac{4\epsilon^2}{b_0^2 (1-\epsilon^2)} e^{-\frac{c}{2}\Delta \chi (2-\frac{2}{\epsilon})} \bar{t}^{2-\frac{2}{\epsilon}} + \cdots \right)\,,
\end{align}
which, using the transformation of the integration constant $\bar{b}_0^2 = b_0^2 e^{\frac{\epsilon - 1}{\epsilon} c\Delta \chi}$ \cite{Lehners:2015sia}, leads to
\begin{align}
\bar{b}_\pm (\bar{t})= b_{\infty \pm} \left(1 + \frac{4\epsilon^2}{\bar{b}_0^2 (1-\epsilon^2)} \bar{t}^{2-\frac{2}{\epsilon}} + \cdots \right)\,.
\end{align}
Thus the solution for the anisotropy parameters is indeed unchanged in form, and in particular the value of the anisotropies at infinity is unchanged.

We are now in a position to determine how the action changes along a classical history. As argued above, at sufficiently late times the anisotropies will act as small perturbations, and hence we can treat them perturbatively without loss of generality. Then, to leading order, the $b_\pm$ dependent changes in the action \eqref{onshell} will be reflected solely in the term
\begin{equation}
\int \mathrm{d}t N a \,U(\beta_+, \beta_-)\,.
\end{equation}
Successive instantons are obtained in the late time limit by evolving in the Lorentzian time direction, hence the lapse function is $N=1$ and the asymptotic scaling of the anisotropy parameters in Eq. \eqref{AsymptoticBetas} implies that they will reach constant values,
\begin{align}
\Delta Re(S) &= \int \mathrm{d}t a \,U(\beta_+, \beta_-) \\ &\approx \int \mathrm{d}t \,b_0 \,t^{\frac{1}{\epsilon}} U(b_+, b_-) \\ &\approx b_0 \,t^{\frac{1}{\epsilon}+1} U \\ &\propto b\,V^{-1/2} U\,.
\end{align}
Thus $\partial_{b_\pm} Re(S) \propto b\,V^{-1/2} U_{,b_\pm}.$ In order to determine the change in the imaginary part of the action, we can use the scaling/shift symmetry described above. As shown in \cite{Lehners:2015sia}, for isotropic instantons with constant $\epsilon$ this symmetry implies that $\Delta Im(S) \propto b^{\frac{2\epsilon}{\epsilon - 1}}V^{\frac{1}{\epsilon - 1}}.$ But we have just seen that the symmetry does not affect the anisotropy parameters. Hence we must have 
\begin{align}
\Delta Im(S) \propto f(b_+,b_-) \,b^{\frac{2\epsilon}{\epsilon - 1}}V^{\frac{1}{\epsilon - 1}}\,,
\end{align}
for some function $f(b_+,b_-)$ which we cannot determine from these arguments. However, we also do not need to know its precise functional form. This is because asymptotically, both the function $f$ and its derivatives $f_{,b_\pm}$ will reach the constant values $f(b_{\pm\infty})$ and $f_{,b_\pm}(b_{\pm\infty})$ respectively, and we are only interested in the overall scaling. Putting the above results together, we arrive at the scaling law for the WKB classicality conditions associated with the anisotropy parameters,
\begin{align}
{\cal{WKB}}_{\, b_\pm} = \frac{\partial_{b_\pm} Im(S)}{\partial_{b_\pm} Re(S)} \propto \frac{f_{,b_\pm}(b_{\pm\infty})\, b^{\frac{2\epsilon}{\epsilon - 1}}V^{\frac{1}{\epsilon - 1}}}{U_{,b_\pm}(b_{\pm\infty})b\,V^{-1/2}} \propto \frac{1}{b^{1+\epsilon}}\,.
\end{align}

\begin{figure}[h] 
\begin{center}
\includegraphics[width=0.75\textwidth]{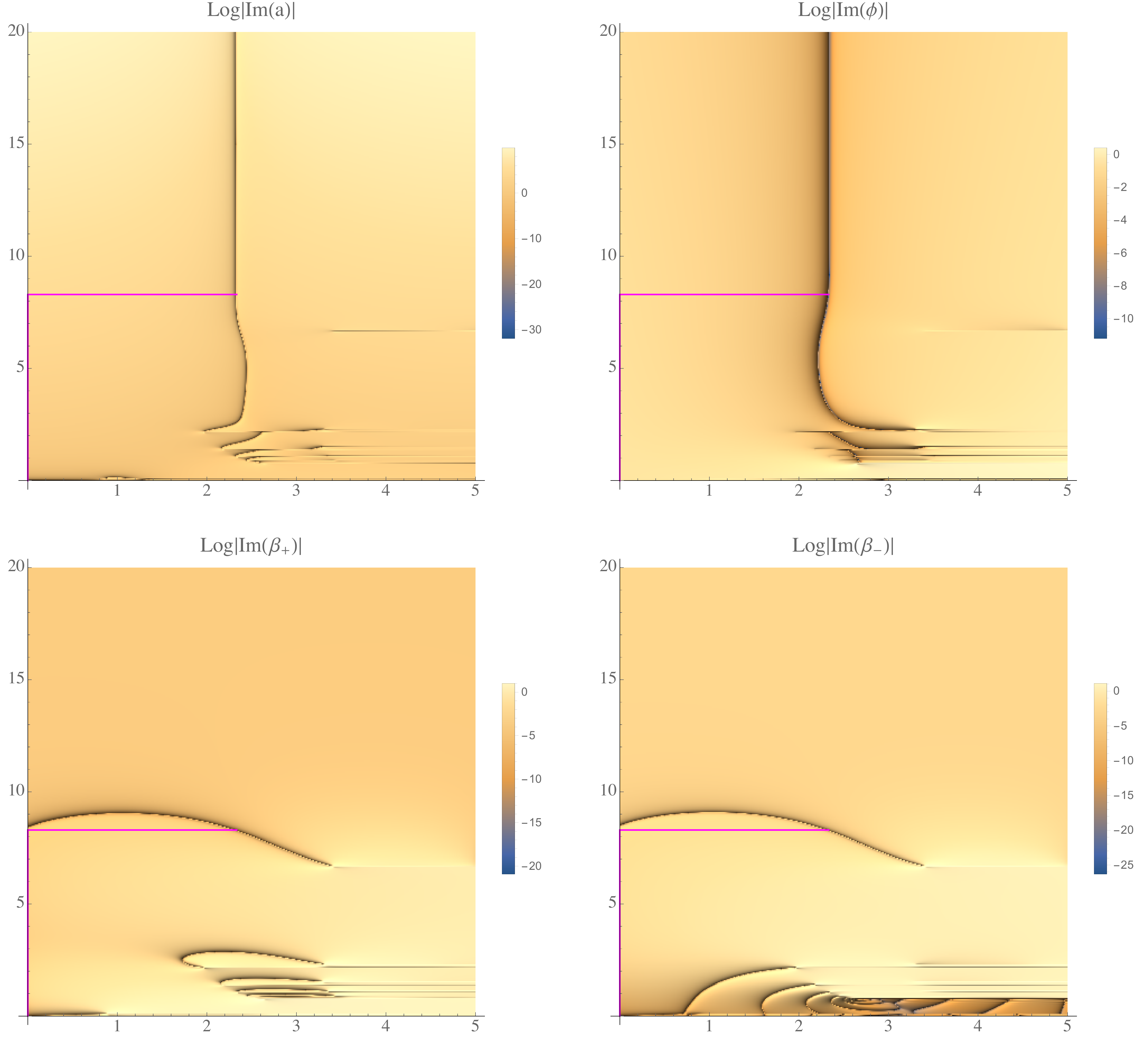}
\caption{An ``early'' instanton with a smaller scale factor, optimised for $b=100, \chi=-1/2, b_+=1, b_-=1.$ These values are reached at $\tau_f = 2.33345 + 8.29691\,i$, with the South Pole values $\phi_{SP}=0.905134 - 0.554599\,i, \beta_{SP+}''=
-0.909196 + 0.164990\,i, \beta_{SP-}'' =
-0.00369960 + 0.000726549\,i$. The magenta contour runs from the South Pole at $\tau=0$ to the final hypersurface at $\tau_f.$ We have solved the equations of motion over a larger time domain in order to show that the fields (especially the anisotropy functions) do not retain approximately real values beyond $\tau_f$ yet.}
\label{fig:beforedip}
\end{center}
\end{figure}

A final feature seen in Fig. \ref{fig:wkb} is the little dip in the plot of ${\cal{WKB}}_{\,b}$. This feature shows that the scaling law has not been reached yet, and thus suggests that the wavefunction has not really reached classicality yet at this stage. It is instructive to look at an early instanton just before the dip -- such an instanton is shown in Fig. \ref{fig:beforedip}. The instanton has been optimised to reach the values $(b=100, \chi= -1/2, b_+ = 1, b_- = 1).$ Interestingly, the vertical lines emanating from $\tau_f$ for the plots of the imaginary values of the scale factor and scalar field show that these fields are already very nearly real in the Lorentzian time direction, while the ansiotropy parameters do not remain as close to real beyond $\tau_f,$ compare also to Fig. \ref{fig:afterdip} This is in agreement with the fact that the classicality conditions involving the anisotropy functions are satisfied more slowly than those involving the scale factor and scalar field. Thus, at that stage, one cannot yet say that a classical spacetime is predicted, and several more e-folds of expansion are needed before classicality is reached.

\section{Discussion}

We have shown that anisotropic (Bianchi IX) no-boundary inflationary instantons may be constructed with arbitrary values of the anisotropy functions. A novel feature is that the construction of these instantons requires a different contour in the complex time plane than the one usually employed for no-boundary inflationary instantons, due to the presence of singularities caused by the anisotropies. A further implication of the anisotropies is that the wavefunction of the universe becomes classical in a WKB sense less fast than in the isotropic case. More precisely, the classicality conditions are satisfied only in inverse proportion to the linear size of the universe, as opposed to inversely to the volume, which would have been the case for isotropic instantons. Thus the anisotropies keep the wavefunction fully quantum for longer, and it will be interesting to explore possible implications of this feature.

Our results imply that for a scalar field model with an inflationary potential the no-boundary state predicts classical histories with arbitrarily large anisotropies. In all cases that we have constructed we found that an inflationary phase is reached, and thus at late times these anisotropies decay away. In general, due to the presence of the anisotropies, the classical histories reached at late times  contain a big bang singularity when extrapolated into their past. This singularity is then resolved by the no-boundary proposal in the sense that the description in terms of a classical spacetime becomes untenable at small scale factor values, since the wavefunction does not yet describe a classical universe at that point. 

There are several avenues for future work: we have not much discussed the relative probabilities of obtaining different classical histories. This is because our results lead to a puzzle that will require a more detailed investigation. Indeed, as one can see from Fig. \ref{fig:action} the imaginary part of the action passes through zero and changes sign. This is puzzling from the point of view of a Picard-Lefschetz analysis of the path integral, since this analysis suggests that the relevant saddle points of the path integral must always be smaller in magnitude than $1,$ see the discussion in \cite{FLT}. But this would mean that when the imaginary part of the action crosses zero one must switch to using a different, complex conjugate, saddle point. In turn such a switch would imply that the peak of the wavefunction would not reside at zero anisotropy, as currently assumed in the literature, but at the values of the anisotropies where the imaginary part of the action vanishes. However, performing a full Picard-Lefschetz analysis of the Bianchi IX model is beyond the scope of the present work, and we hope to return to this issue in the near future. What can be said either way is that the solutions with large anisotropies and correspondingly large imaginary part of the action lead to a highly suppressed wavefunction.

A further extension concerns the construction of anisotropic ekpyrotic instantons. Here also we are naively faced with a puzzle: the new ${\cal{WKB}}_{\,b_\pm}$ classicality conditions that we have derived here scale as $b^{-1-\epsilon}.$ In ekpyrotic models the universe is contracting and moreover $\epsilon > 3.$ Then, if the same scaling were to hold, it would appear that the classicality conditions would blow up and not be satisfied as $b$ shrinks. This is however hard to believe as an ekpyrotic phase is an attractor and suppresses anisotropies in much the same way as inflation does. It will therefore be interesting to clarify this puzzle.

\acknowledgments

The work of SFB is supported in part by a grant from the Studienstiftung des Deutschen Volkes.

\bibliographystyle{utphys}
\bibliography{anisotropy}

\providecommand{\href}[2]{#2}\begingroup\raggedright\begin{thebibliography}{10}

\bibitem{Baumann:2009ds}
D.~Baumann,
  \href{http://dx.doi.org/10.1142/9789814327183_0010}{``{Inflation},''} in {\em
  {Physics of the large and the small, TASI 09, proceedings of the Theoretical
  Advanced Study Institute in Elementary Particle Physics, Boulder, Colorado,
  USA, 1-26 June 2009}}, pp.~523--686.
\newblock 2011.
\newblock
\href{http://arxiv.org/abs/0907.5424}{{\tt arXiv:0907.5424 [hep-th]}}.
\newblock

\bibitem{Lehners:2008vx}
J.-L. Lehners, ``{Ekpyrotic and Cyclic Cosmology},''
  \href{http://dx.doi.org/10.1016/j.physrep.2008.06.001}{{\em Phys. Rept.} {\bf
  465} (2008)  223--263},
\href{http://arxiv.org/abs/0806.1245}{{\tt arXiv:0806.1245 [astro-ph]}}.

\bibitem{Hawking:1981gb}
S.~W. Hawking, ``{The Boundary Conditions of the Universe},''
{\em Pontif. Acad. Sci. Scr. Varia} {\bf 48} (1982)  563--574.

\bibitem{Hartle:1983ai}
J.~B. Hartle and S.~W. Hawking, ``{Wave Function of the Universe},''
\href{http://dx.doi.org/10.1103/PhysRevD.28.2960}{{\em Phys. Rev.} {\bf D28}
  (1983)  2960--2975}.

\bibitem{Hawking:1983hj}
S.~W. Hawking, ``{The Quantum State of the Universe},''
\href{http://dx.doi.org/10.1016/0550-3213(84)90093-2}{{\em Nucl. Phys.} {\bf
  B239} (1984)  257}.

\bibitem{Hartle:2008ng}
J.~B. Hartle, S.~W. Hawking, and T.~Hertog, ``{The Classical Universes of the
  No-Boundary Quantum State},''
  \href{http://dx.doi.org/10.1103/PhysRevD.77.123537}{{\em Phys. Rev.} {\bf
  D77} (2008)  123537},
\href{http://arxiv.org/abs/0803.1663}{{\tt arXiv:0803.1663 [hep-th]}}.

\bibitem{Vilenkin:1982de}
A.~Vilenkin, ``{Creation of Universes from Nothing},''
\href{http://dx.doi.org/10.1016/0370-2693(82)90866-8}{{\em Phys. Lett.} {\bf
  B117} (1982)  25--28}.

\bibitem{Vilenkin:1983xq}
A.~Vilenkin, ``{The Birth of Inflationary Universes},''
\href{http://dx.doi.org/10.1103/PhysRevD.27.2848}{{\em Phys. Rev.} {\bf D27}
  (1983)  2848}.

\bibitem{Vilenkin:1984wp}
A.~Vilenkin, ``{Quantum Creation of Universes},''
\href{http://dx.doi.org/10.1103/PhysRevD.30.509}{{\em Phys. Rev.} {\bf D30}
  (1984)  509--511}.

\bibitem{Vilenkin:1986cy}
A.~Vilenkin, ``{Boundary Conditions in Quantum Cosmology},''
\href{http://dx.doi.org/10.1103/PhysRevD.33.3560}{{\em Phys. Rev.} {\bf D33}
  (1986)  3560}.

\bibitem{Belinsky:1970ew}
V.~A. Belinsky, I.~M. Khalatnikov, and E.~M. Lifshitz, ``{Oscillatory approach
  to a singular point in the relativistic cosmology},''
\href{http://dx.doi.org/10.1080/00018737000101171}{{\em Adv. Phys.} {\bf 19}
  (1970)  525--573}.

\bibitem{Hawking:1984wn}
S.~W. Hawking and J.~C. Luttrell, ``{The Isotropy of the Universe},''
\href{http://dx.doi.org/10.1016/0370-2693(84)90809-8}{{\em Phys. Lett.} {\bf
  B143} (1984)  83}.

\bibitem{Wright:1984wm}
W.~A. Wright and I.~G. Moss, ``{The Anisotropy of the Universe},''
\href{http://dx.doi.org/10.1016/0370-2693(85)90569-6}{{\em Phys. Lett.} {\bf
  B154} (1985)  115--119}.

\bibitem{delCampo:1989hy}
S.~del Campo and A.~Vilenkin, ``{Tunneling Wave Function for Anisotropic
  Universe},''
\href{http://dx.doi.org/10.1016/0370-2693(89)91047-2}{{\em Phys. Lett.} {\bf
  B224} (1989)  45--48}.

\bibitem{Amsterdamski:1985qu}
P.~Amsterdamski, ``{Wave Function of an Anisotropic Universe},''
\href{http://dx.doi.org/10.1103/PhysRevD.31.3073}{{\em Phys. Rev.} {\bf D31}
  (1985)  3073--3078}.

\bibitem{Duncan:1988zq}
M.~J. Duncan and L.~G. Jensen, ``{The Quantum Cosmology of an Anisotropic
  Universe},''
\href{http://dx.doi.org/10.1016/0550-3213(89)90576-2}{{\em Nucl. Phys.} {\bf
  B312} (1989)  662--672}.

\bibitem{Fujio:2009my}
K.~Fujio and T.~Futamase, ``{Appearance of classical Mixmaster Universe from
  the No-Boundary Quantum State},''
  \href{http://dx.doi.org/10.1103/PhysRevD.80.023504}{{\em Phys. Rev.} {\bf
  D80} (2009)  023504},
\href{http://arxiv.org/abs/0906.2616}{{\tt arXiv:0906.2616 [gr-qc]}}.

\bibitem{Battarra:2014kga}
L.~Battarra and J.-L. Lehners, ``{On the No-Boundary Proposal for Ekpyrotic and
  Cyclic Cosmologies},''
  \href{http://dx.doi.org/10.1088/1475-7516/2014/12/023}{{\em JCAP} {\bf 1412}
  (2014) no.~12, 023},
\href{http://arxiv.org/abs/1407.4814}{{\tt arXiv:1407.4814 [hep-th]}}.

\bibitem{Battarra:2014xoa}
L.~Battarra and J.-L. Lehners, ``{On the Creation of the Universe via Ekpyrotic
  Instantons},'' \href{http://dx.doi.org/10.1016/j.physletb.2015.01.028}{{\em
  Phys. Lett.} {\bf B742} (2015)  167--171},
\href{http://arxiv.org/abs/1406.5896}{{\tt arXiv:1406.5896 [hep-th]}}.

\bibitem{Bramberger:2016yog}
S.~F. Bramberger, G.~Lavrelashvili, and J.-L. Lehners, ``{Quantum tunneling
  from paths in complex time},''
  \href{http://dx.doi.org/10.1103/PhysRevD.94.064032}{{\em Phys. Rev.} {\bf
  D94} (2016) no.~6, 064032},
\href{http://arxiv.org/abs/1605.02751}{{\tt arXiv:1605.02751 [hep-th]}}.

\bibitem{Misner:1969hg}
C.~W. Misner, ``{Mixmaster universe},''
\href{http://dx.doi.org/10.1103/PhysRevLett.22.1071}{{\em Phys. Rev. Lett.}
  {\bf 22} (1969)  1071--1074}.

\bibitem{Damour:2002et}
T.~Damour, M.~Henneaux, and H.~Nicolai, ``{Cosmological billiards},''
  \href{http://dx.doi.org/10.1088/0264-9381/20/9/201}{{\em Class. Quant. Grav.}
  {\bf 20} (2003)  R145--R200},
\href{http://arxiv.org/abs/hep-th/0212256}{{\tt arXiv:hep-th/0212256
  [hep-th]}}.

\bibitem{Kleinschmidt:2009cv}
A.~Kleinschmidt, M.~Koehn, and H.~Nicolai, ``{Supersymmetric quantum
  cosmological billiards},''
  \href{http://dx.doi.org/10.1103/PhysRevD.80.061701}{{\em Phys. Rev.} {\bf
  D80} (2009)  061701},
\href{http://arxiv.org/abs/0907.3048}{{\tt arXiv:0907.3048 [gr-qc]}}.

\bibitem{Koehn:2011ph}
M.~Koehn, ``{Relativistic Wavepackets in Classically Chaotic Quantum
  Cosmological Billiards},''
  \href{http://dx.doi.org/10.1103/PhysRevD.85.063501}{{\em Phys. Rev.} {\bf
  D85} (2012)  063501},
\href{http://arxiv.org/abs/1107.6023}{{\tt arXiv:1107.6023 [gr-qc]}}.

\bibitem{Halliwell:1988wc}
J.~J. Halliwell, ``{Derivation of the Wheeler-De Witt Equation from a Path
  Integral for Minisuperspace Models},''
\href{http://dx.doi.org/10.1103/PhysRevD.38.2468}{{\em Phys. Rev.} {\bf D38}
  (1988)  2468}.

\bibitem{Vilenkin:1988yd}
A.~Vilenkin, ``{The Interpretation of the Wave Function of the Universe},''
\href{http://dx.doi.org/10.1103/PhysRevD.39.1116}{{\em Phys. Rev.} {\bf D39}
  (1989)  1116}.

\bibitem{Lyons:1992ua}
G.~W. Lyons, ``{Complex solutions for the scalar field model of the
  universe},''
\href{http://dx.doi.org/10.1103/PhysRevD.46.1546}{{\em Phys. Rev.} {\bf D46}
  (1992)  1546--1550}.

\bibitem{Bramberger:2017cgf}
S.~F. Bramberger, T.~Hertog, J.-L. Lehners, and Y.~Vreys, ``{Quantum
  Transitions Through Cosmological Singularities},''
\href{http://arxiv.org/abs/1701.05399}{{\tt arXiv:1701.05399 [hep-th]}}.

\bibitem{Lehners:2015sia}
J.-L. Lehners, ``{Classical Inflationary and Ekpyrotic Universes in the
  No-Boundary Wavefunction},''
  \href{http://dx.doi.org/10.1103/PhysRevD.91.083525}{{\em Phys. Rev.} {\bf
  D91} (2015) no.~8, 083525},
\href{http://arxiv.org/abs/1502.00629}{{\tt arXiv:1502.00629 [hep-th]}}.

\bibitem{FLT}
J.~Feldbrugge, J.-L. Lehners, and N.~Turok {\em in preparation} (2017)  .

\end{thebibliography}\endgroup

\end{document}